\documentclass[twocolumn, twocolappendix, trackchanges]{aastex701}
\usepackage{amsmath} 
\usepackage{physics}
\usepackage{enumitem}

\begin{document}

\title{Resonance and Stochastic Dynamics of Interplanetary Dust}

\author[orcid=0009-0000-9655-3029,sname='Qiu']{Minli Qiu}
\affiliation{Department of Astronomy, University of Virginia, Charlottesville, VA 22904, USA}
\email[show]{mku9uy@virginia.edu}

\author[orcid=0000-0001-5611-1349,sname='Arras']{Phil Arras}
\affiliation{Department of Astronomy, University of Virginia, Charlottesville, VA 22904, USA}
\affiliation{Virginia Institute for Theoretical Astronomy, University of Virginia, Charlottesville, VA 22904, USA.}
\email{arras@virginia.edu}

\begin{abstract}

We study the motion of dust particles inspiraling from distant dust reservoirs toward a close-in planetary system, including the combined effects of radiation pressure and Poynting-Robertson (PR) drag. As dust particles migrate inward, they can be trapped in mean motion resonances (MMRs) depending on the competition between the planet's gravity and PR drag. Our goals are to understand the conditions under which particles can be trapped in MMRs, the evolution in and eventual escape from resonance, and the fraction of dust particles which will hit the planetary upper atmosphere, seeding it with heavy elements. Low-order eccentricity expansions of the disturbing function break down, so we employ an exact, rapid-phase--averaged disturbing function to determine resonant equilibrium points and their stability. We derive analytic expressions for the growth rate of dissipative equilibrium points and confirm that librations are generically overstable in important low-order resonances such as $3\!:\!1$, $2\!:\!1$, and $3\!:\!2$, implying that resonant capture is ultimately temporary and dust particles escape over a wide range of planet masses and dust size. After resonance escape, the dust orbit is planet-crossing and its subsequent evolution is intrinsically stochastic, governed by repeated close encounters that produce random gravitational kicks. We develop an analytic, epicycle-based scattering model to derive the impact parameter distribution $P(b)$ and the resulting energy change distribution $P(\Delta x)$, including a transition of the power-law tail from $|\Delta x|^{-2}$ in coplanar encounters to $|\Delta x|^{-3}$ at large inclination. Using these encounter distributions, we construct a Monte Carlo method that predicts the fractions of dust particles that collide with the planet, sublimate near the star, or are ejected. Comparison of the Monte Carlo calculations with orbit integrations shows good agreement across the cases studied.
\end{abstract}

\keywords{
\href{https://astrothesaurus.org/uat/821}{Interplanetary dust (821)} --
\href{https://astrothesaurus.org/uat/500}{Exozodiacal dust (500)} --
\href{https://astrothesaurus.org/uat/487}{Exoplanet atmospheres (487)} --
\href{https://astrothesaurus.org/uat/211}{Celestial mechanics (211)} --
\href{https://astrothesaurus.org/uat/1181}{Orbital resonances (1181)} --
\href{https://astrothesaurus.org/uat/255}{Close encounters (255)}
}

\section{Introduction}\label{sec:intro}

Dust in the Solar System is continually generated by asteroidal and cometary sources, and gradually spirals inward towards the Sun due to Poynting-Robertson (PR) drag \citep{2013pss3.book..431M}. As the dust migrates inward, it can undergo several fates, including physical collision with planets, ejection from the Solar System, or further inspiraling towards interior planets or the Sun.

Our study is motivated by the possibility that accretion of
interplanetary dust may enrich the upper atmospheres of close-in
exoplanets to levels detectable in transmission spectra and interpreted as ``high-altitude clouds". Recent detailed studies by \citet{2017ApJ...847...32L, 2022ApJ...932...90A, 2022ApJ...927..184M, 2024ApJ...974..190M, Chang2026Dust} have explored the affect of opacity from accreted solids and ablated atoms/molecules on emission and transmission spectra. The key parameter is the accretion rate of dust into the planet's upper atmosphere. This quantity can be expressed as the product of the local amount of orbiting dust, set by PR-driven accretion of interplanetary dust inward toward the star, and the efficiency at which the planet sweeps up the orbiting grains. In this paper we focus on the orbital dynamics problem underlying the accretion efficiency. It is a ``migration into resonance" problem, similar to that studied for planet pairs interacting with a gaseous disk, or satellite pairs subject to tidal friction torques.

\citet{2018MNRAS.480.5560B} computed dust accretion rates across a range of exoplanetary systems and found that close-in, massive planets are most efficient at accreting dust. Similarly,
\citet{2022ApJ...932...90A} have shown that the accretion rate of interplanetary dust onto close-in exoplanets may be much larger than that onto the planets in our own Solar System. This occurs for two main reasons. First, a large fraction of the interplanetary dust may be accreted into the atmospheres of close-in planets, where collisions are favored over ejections. In the extreme limit of 100\% accretion efficiency and ignoring destructive collisions between the reservoir and the planet, all the dust created by the outer dust reservoirs is accreted into the planet's upper atmosphere, after which it sediments downward through the region where the transmission spectrum is formed. Secondly, hot dust close to the star is quite common in nearby stars, with 15–20\% showing detectable hot dust excess \citep{2013A&A...555A.104A, 2014A&A...570A.128E, 2021A&A...651A..45A}. Detectability against the glare of the host star, however, requires an enormous factor $\gtrsim 10^3$ more dust that in our own solar system. \citet{2022ApJ...932...90A} show that even undetectable levels of hot dust, $1-10^3$ times that in our solar system, may be able to significantly enrich the upper atmospheres of close-in planets. Hence they suggest the possibility that the upper atmosphere clouds seen in exoplanet transmission spectra may possibly be due to accretion of interplanetary dust.

Numerous studies have investigated the dynamical effects of PR drag on dust particles in various contexts. Early work by
\citet{1979Icar...40....1B} established the dependence of radiation forces on
particle size and composition through the radiation pressure efficiency factor
$\beta$. A particularly important problem is the competition between PR drag and gravitational perturbations from a planet, which can lead to temporary capture of migrating dust particles into mean motion resonances
\citep{1975Icar...25..489G}. \cite{1993Icar..104..244W} examined conditions for capture into resonance for the circular restricted three-body problem and derived constraints on $\beta$, dust particle eccentricity, and planetary mass. They also analytically derived the dust particle's eccentricity evolution inside resonance, but as we will show, the eccentricity and resonant angle show key differences compared to exact solutions. In these cases, dust particles are captured into ``asymmetric resonances", where the resonant angle $\phi_2$ librates away from the apsidal axis instead of $\phi \sim 0$ or $\phi \sim \pi$ \citep{2005ApJ...619..623M}. Using a resonance-averaged disturbing function for the gravitational perturbation of the planet, \cite{1994Icar..110..239B} identified those off-axis fixed points and computed particle trajectories in the phase space. After exploring the linear stability of the off-axis fixed points, for planet masses $\in [10^{-7}, 10^{-3}]M_{\odot}$ and for $\beta \in [0, 0.5]$, all the solutions were found to be unstable. This implies that the dust particle would always get out of resonance after a certain time interval. Numerical work by \cite{1994A&A...289..972S} reached a similar conclusion, indicating that small oscillations around the equilibrium point must grow exponentially with time and no permanent capture may be expected.

Previous studies on the stability of equilibrium points have relied largely on numerical methods. Here we derive analytic expressions for the
growth/damping rates of these equilibrium points, providing a physical explanation for the instabilities that ultimately cause dust particles to escape resonance. We also examine the role of higher-order terms in the traditional Laplace-coefficient expansion of the disturbing function, especially at high equilibrium eccentricities. These higher-order terms become essential as lower-order approximations lose accuracy, particularly in cases of $2\!:\!1$ and $3\!:\!1$ MMRs, where stability conditions are especially sensitive.

The resonance escape mechanism of dust particles shares similarities with overstable librations in multi-planetary systems \citep{2014AJ....147...32G}. Their work demonstrated that eccentricity damping plays a crucial role in inducing overstability in librations during convergent migration, resulting in temporary resonance capture governed by the equilibrium eccentricity relative to the planet-to-star mass ratio. Similarly, \citet{2015ApJ...810..119D} showed that resonance stability is highly sensitive to the mass ratio and the eccentricity damping timescales of the interacting bodies. These studies show how resonance stability connects planet migration and dust evolution, revealing shared dynamics from small particles to planets under dissipative forces.

Once the dust particle escapes from resonance, its subsequent evolution becomes intrinsically stochastic and is no longer amenable to analytic treatment using perturbation theory.  
We therefore model the post-resonant evolution as a diffusion process governed by probabilistic scattering statistics and solve it using a Monte Carlo approach. Extensive numerical studies of cometary diffusion have been carried out in this context. A standard approach is to record the change in orbital energy ($\Delta x$) at each periapsis passage for a large ensemble of randomly distributed orbits and to characterize the resulting statistics \citep{1981A&A....96...26F}. Numerical experiments by \citet{1968AJ.....73.1039E} and
\citet{1987AJ.....94.1330D} showed that the distribution of orbital energy changes during planetary encounters exhibits an approximate power-law tail. Here we derive this distribution analytically using an epicycle model in the rotating frame. We derive the impact parameter distribution for planet–dust encounters analytically and combine it with a gravitational kick model to obtain the corresponding distribution of orbital energy changes. In our model, the resulting distribution exhibits different power-law behaviors at small and large inclinations and smoothly interpolates between these limits at intermediate inclinations, rather than following a single universal power law. The analytic predictions are in good agreement with numerical experiments. In contrast to the classical encounter formalism of \citet{1976iecg.book.....O}, which focuses only on encounters occurring within the Hill sphere, our framework includes encounters both inside and outside the Hill radius and derives the full impact parameter distribution across these regimes. 

While orbit integrations give an accurate incorporation of all physical effects, their expense and complexity motivate simple models to understand the physical processes and dependence on the parameters. For predicting accretion and ejection fractions, previous studies typically assume fixed per-orbit probabilities for both planetary collisions and ejections, ignoring planetary perturbations. The rate of collisions with the sublimation zone of the star is not computed explicitly, but rather are the absence of collisions with the planet or ejection. In particular, \citet{2018MNRAS.480.5560B} fit the results of orbit integrations to a parametrized dependence on planet mass, orbital separation, etc. By contrast, we use analytically derived encounter statistics to compute collision and Hill-entry probabilities, while treating all non-collisional outcomes as arising from dynamical diffusion process rather than assigning an explicit per-orbit ejection probability. These elements are incorporated into a Monte Carlo framework to model the final fates of dust particles. Direct comparison with numerical simulations shows good agreement across all cases studied.

This paper is structured as follows. Section~\hyperref[sec2]{2} establishes a new averaged disturbing function valid for high eccentricities, and derives the equations of motion for dust particles under dissipation. We also explore the evolution of level curves in phase space for different planet masses, tracking the progression of fixed points. In Section \hyperref[sec3]{3}, we present numerical results from orbital integrations, illustrating the orbital elements' evolution across various resonances, with particular focus on asymmetric capture in $2\!:\!1$ resonance. In Section~\hyperref[sec4]{4} we develop semi-analytic fixed-point solutions, analyze their stability, and show that dissipative equilibrium are generically overstable. Section \hyperref[sec5]{5} details numerical integrations over different planet masses and $\beta$ values, establishing criteria for capture into resonance. Section~\hyperref[sec6]{6} describes the stochastic evolution after resonance escape and derives probabilistic outcomes for dust fates through Monte Carlo process. Our main conclusions are summarized in Section~\hyperref[sec7]{7}. Appendix \ref{app:loworderexpansion} contains a detailed comparison of low-order expansions of the disturbing function and Lagrange's planetary equations to orbital integrations, illustrating the pitfalls of such expansions. Appendix \ref{app:impact_parameter} contains a derivation of the toy epicycle model for the impact parameter distribution. Appendix \ref{app:pdx} contains a review of the energy change due to close encounters with the planet, and carries out a change of variables to find the distribution of energy changes from that of impact parameters.

\section{Circular Restricted Three-Body Problem with Radiation Forces}\label{sec2}

We consider dust particles originating in distant reservoirs that migrate
inward toward a close-in planetary system with semi-major axis $a_1 \lesssim  1\, \rm AU$ under the influence of PR drag. For dust particles initially at $a_2 \gg a_1$, PR drag  circularizes their orbit while reducing their semi-major axis. By the time the particles reach the region of the planet, their orbits are therefore nearly circular. As the dust approaches the planet, and for sufficiently large planet mass, the dust
may enter MMRs with the planet, halting the orbital decay. 

Previous numerical studies (e.g., \citealt{1994Icar..110..239B}) have
shown that the resonant fixed points are overstable: small librations around the equilibrium grow with time, eventually causing the particle to escape resonance. Eventually there are three possible outcomes \citep{2018MNRAS.480.5560B}: (1) collision with the planet, depositing dust in the planetary atmosphere; (2) sublimation of the dust particle near the star; or (3) ejection from the system. For planets far from the star, sublimation occurs after the dust particle has detached from the planet. Subsequent PR-driven orbital decay and circularization will lead to a nearly circular orbit when the dust hits the sublimation zone. In our study, we include a finite radius for the dust sublimation zone. This introduces a new channel for planets close to the star in which particles undergo ``orbital diffusion" to high eccentricity and directly hit the sublimation zone while still interacting with the planet.

In this section, we present the equations of motion and their numerical integration, and then study the resonant dynamics that governs capture into and evolution within mean motion resonances.

\subsection{Equations of Motion}\label{sec2.1}

The star (subscript 0) of mass $m_0$ and planet (subscript 1) of mass $m_1$ are in a circular orbit with semi-major axis $a_1$. The dust particle (subscript 2) is initiated on a circular orbit with semi-major axis $a_2$ outside the inner binary. The ratio of the  radiation pressure to gravitational force from the star is given by \citep{1979Icar...40....1B}
\begin{equation}
\beta = \frac{3LQ_{\mathrm{pr}}}{16\pi G m_0 c\rho_d r_d},
\end{equation}
where $L$ is the stellar luminosity, $G$ is the gravitational constant, $c$ is the speed of light, and $\rho_d$, $r_d$ are the density and radius of the dust particle. $Q_{\mathrm{pr}}$ is the dimensionless radiation pressure efficiency factor; for a perfectly absorbing particle much larger than the wavelength $Q_{\mathrm{pr}} = 1$. In general $Q_{\rm pr}$ depends on the stellar spectrum, as well as the size, shape, and composition of the dust.

For the dust particle, we take gravitational forces from both the star and the planet into account, along with the radiation pressure and PR drag forces from the star. These forces can be expressed as \citep{1979Icar...40....1B}
\begin{equation}
\begin{split}
\vb*{F} &= \vb*{F}_{\mathrm{grav}} + \vb*{F}_{\mathrm{rad}}\\
&= \frac{Gm_0m_2}{r_{02}^3}{\vb*r}_{20} + \frac{Gm_1m_2}{r_{12}^3}{\vb*r}_{21}\\
& \quad + \beta\frac{Gm_0m_2}{{r_{02}}^2}\left[\left(1-\frac{\dot{r}_{02}}{c}\right)\vb*{n}_{02} - \frac{\vb*{v}_{02}}{c}\right], \\
\end{split}
\end{equation}
where $\vb*x_0, \vb*x_1, \vb*x_2$ denote the positions of the star, planet, and dust particle, respectively.
Here $\vb*{r}_{02} = \vb*x_2 - \vb*x_0 = r_{02}\vb*{n}_{02} = -\vb*{r}_{20} $ and $\vb*{r}_{12} = \vb*x_2 - \vb*x_1 = -\vb*{r}_{21}$ represent the star-dust separation vector and the planet-dust separation vector, and $\vb*{v}_{02}$ is the relative velocity of the dust particle with respect to the star. Outward radiation pressure and inward gravity partially cancel, giving the star an effective mass $m_0(1-\beta)$.

The exact equations of motion (EEOM) of the dust particle, star, and planet are integrated using the \texttt{REBOUND} code \citep{2012A&A...537A.128R}, with radiation pressure and PR drag forces incorporated via \texttt{REBOUNDx} \citep{2020MNRAS.491.2885T}. While the star-planet binary is on a known circular orbit, unaffected by the dust (treated as massless), we nevertheless find it convenient to use \texttt{REBOUND} to integrate the equations of motion of all three particles due to the accuracy of the \texttt{IAS15} integrator, \texttt{REBOUND}'s ability to detect collisions, and the convenience in converting Cartesian coordinates to orbital elements. We adopt the planet mass–radius relation $R_1=R_1(m_1)$ from \citet{2017ApJ...834...17C}
to estimate the planetary radius for a given mass. A dust particle is
considered to collide with the planet when the planet–dust separation
becomes smaller than $R_1$, at which point it is removed
from the simulation. We compute the dust sublimation radius $R_{\rm sub}$ around the star by equating the dust equilibrium temperature to the silicate melting temperature, $T \simeq 1600\,\mathrm{K}$ \citep{2011piim.book.....D}, and solving for the corresponding distance from the star. When the star–dust separation falls inside this distance, the particle is assumed to sublimate and is likewise removed from the simulation. Moreover, we monitor the particle’s orbital energy with respect to the star, and classify the particle as ejected when this energy becomes positive and also the particle is outside the critical separation $10a_1$, to make sure the planet's effect is negligible. Ejected particles are also removed from the simulation.

\subsection{Resonance Dynamics}

In this section we discuss an approximation to the dynamics, including only effects from a particular $j:j-k$ exterior
mean motion resonance. The goal is to understand stable equilibrium points which the trajectory may follow. We will focus on coplanar dust and planet orbits for simplicity. The orbital elements used are the semi-major axis $a_2$, eccentricity $e_2$, argument of pericenter $\varpi_2$, and mean longitude $\lambda_2$. The orbital elements are expressed in the heliocentric frame, to take advantage of the analytic disturbing function expansion in \citet{1999ssd..book.....M}. The subscripts 1 and 2 refer to the orbit of the planet and dust around the star, respectively. The mean motions are $n_1 = \sqrt{G(m_0+m_1)/a_1^3}$ and $n_2=\sqrt{Gm_0(1-\beta)/a_2^3}$, giving the resonant semi-major axis
\begin{equation}
a_{2,\mathrm{res}} = a_1\left(\frac{j}{j-k}\right)^{2/3}\left(\frac{m_0(1-\beta)}{m_0+m_1}\right)^{1/3}.
\end{equation}
The potential of the planet evaluated at the position of the dust particle is $-R_2$, where
\begin{eqnarray}
&& R_2(a_1,e_1,\varpi_1,\lambda_1,a_2,e_2,\varpi_2,\lambda_2) \nonumber \\ & = & 
\frac{Gm_1}{\left| \vb*{x}_{\rm kep}(a_2,e_2,\varpi_2,\lambda_2) - \vb*{x}_{\rm kep}(a_1,e_1,\varpi_1,\lambda_1) \right|}
\nonumber \\ & - & 
 Gm_1 \frac{\vb*{x}_{\rm kep}(a_1,e_1,\varpi_1,\lambda_1) \cdot \vb*{x}_{\rm kep}(a_2,e_2,\varpi_2,\lambda_2)}{
\left|\vb*{x}_{\rm kep}(a_1,e_1,\varpi_1,\lambda_1) \right|^3 }.
\label{eq:R2}
\end{eqnarray}
Here $\vb*{x}_{\rm kep}$ is the formula for the Cartesian coordinates in terms of the orbital elements, 
and the ``dipole" term in Equation \eqref{eq:R2} arises from the non-inertial reference frame centered on the star.
Lagrange's planetary equations for the orbital elements \citep{1999ssd..book.....M}, including the orbit-averaged dissipative terms from PR drag, are then
\begin{equation}\label{eq:dadt}
\frac{d a_2}{dt} = \frac{2}{n_2a_2}\frac{\partial R_2}{\partial \lambda_2} -\frac{2a_2}{3\tau_{n_2}}\frac{1+\frac{3}{2}e_2^2}{(1-{e_2}^2)^{3/2}},
\end{equation}
\begin{equation}\label{eq:dedt}
\begin{split}
\frac{d e_2}{dt} & = -\frac{\sqrt{1-e_2^2}}{n_2a_2^2e_2}\left(1-\sqrt{1-e_2^2}\right)\frac{\partial R_2}{\partial \lambda_2} - \frac{\sqrt{1-e_2^2}}{n_2a_2^2e_2}\frac{\partial R_2}{\partial \varpi_2}\\
& \quad -\frac{e_2}{\tau_{e_2}}\frac{1}{(1-{e_2}^2)^{1/2}},
\end{split}
\end{equation}
\begin{equation}\label{eq:dpomegadt}
\begin{split}
\frac{d\varpi_2}{dt} = \frac{\sqrt{1-e_2^2}}{n_2a_2^2e_2}\frac{\partial R_2}{\partial e_2},
\end{split}
\end{equation}
\begin{equation}\label{eq:dlambdadt}
\frac{d\lambda_2}{dt} = n_2 -\frac{2}{n_2a_2}\frac{\partial R_2}{\partial a_2} + \frac{\sqrt{1-e_2^2}(1-\sqrt{1-e_2^2})}{n_2a_2^2e_2}\frac{\partial R_2}{\partial e_2}.
\end{equation}
The conservative equations of motion have been augmented by the orbit-averaged changes in $a_2$ and $e_2$ \citep{1950ApJ...111..134W}, where
$\tau_{\rm n2} = (1/3) (a_2^2c)/(Gm_0 \beta)$ and $\tau_{\rm e2} = (2/5) a_2^2c/(Gm_0 \beta)$ are the migration and eccentricity damping timescales at low eccentricity. 

A key difference between the migration and eccentricity damping timescales for PR drag and the more familiar gas disk migration case is that $\tau_{\rm n2} \sim \tau_{\rm e2}$ for PR drag while $\tau_{\rm n2} \gg  \tau_{\rm e2}$ for gas drag \citep{2014AJ....147...32G}. As a result, the dissipative equilibrium points have order unity eccentricity for PR drag, while the eccentricity in the gas drag case is small.

While the above equations for the orbital elements are exact, they include all possible pieces of the disturbing function, e.g., resonant terms, secular terms, short period terms, etc. To make semi-analytic progress, we will focus on a particular $j:j-k$ resonance. However, the large eccentricities of interest imply that leading order expansions of $R_2 \propto e_2^k$ for small $k$ will fail. Previous treatments of resonant dynamics \citep{1989A&A...225..541F} address this issue by analytically averaging $R_2$ after expanding it to finite order in eccentricity. While this captures higher-order structure beyond leading-order theory, the truncation limits accuracy at large eccentricities (e.g., $e_2 \gtrsim 0.6627$). Instead, we choose to numerically average $R_2$ (Equation \ref{eq:R2}) over a rapid phase variable, effectively including all orders in the expansion, up to the integration accuracy. This approach avoids truncation of $R_2$ and preserves its full dependence on orbital geometry, making it well suited for high-eccentricity resonant dynamics.

To perform the averaging over the rapid phase, we follow \cite{1989A&A...225..541F} to trade the variable set $(\lambda_1,\varpi_2,\lambda_2)$ for the inner and outer resonant angles
\begin{equation}
\phi_1 = \frac{1}{k} \left( j\lambda_2 - (j-k)\lambda_1 - k\varpi_1 \right),
\end{equation}
\begin{equation}
\phi_2 = \frac{1}{k} \left( j\lambda_2 - (j-k)\lambda_1 - k\varpi_2 \right),
\end{equation}
and the rapidly varying phase
\begin{equation}
Q = \frac{\lambda_2-\lambda_1}{k},
\end{equation}
as well as $\varpi_1$. The $Q$-averaged disturbing function is then defined as
\begin{eqnarray}\label{eq:Q_ave_R}
&& \langle R_2 \left(a_1,e_1,\phi_1,\varpi_1,a_2,e_2,\phi_2 \right)  \rangle 
\nonumber \\ & = & \int_0^{2\pi} \frac{dQ}{2\pi} R_2(a_1,e_1,\varpi_1,\lambda_1,a_2,e_2,\varpi_2,\lambda_2)
\end{eqnarray}
where the values of $(\varpi_1,\lambda_1,\varpi_2,\lambda_2)$ are determined from $(\phi_1,\phi_2,Q,\varpi_1)$.
We typically use 128 points to perform this $Q$-integral. To compute the derivatives of $\langle R_2 \rangle$ in the planetary equations, we analytically perform the derivatives of the integrand with respect to $(a_2,e_2,\varpi_2,\lambda_2)$ and integrate
these four quantities along with $R_2$ for a total of 5 integrals.

A simplification occurs for a circular inner orbit ($e_1=0$) for which $\langle R_2 \left(a_1,e_1,\phi_1,\varpi_1,a_2,e_2,\phi_2 \right) \rangle$ is independent of $\phi_1$ and $\varpi_1$ \citep{1989A&A...225..541F}. This is the case we explore in this paper. Thus $a_1$ is a fixed parameter and the variable set for the resonant dynamics is $(a_2,e_2,\phi_2)$. The planetary equation for $\phi_2$ is 
\begin{align}\label{eq:dphidt}
\frac{d\phi_2}{dt}
&= \frac{1}{k} \left( j n_2 - (j-k)n_1 \right) \nonumber\\
&\quad + \frac{j}{k} \left(
-\frac{2}{n_2 a_2}\frac{\partial R_2}{\partial a_2}
+ \frac{\sqrt{1-e_2^2}\left(1-\sqrt{1-e_2^2}\right)}{n_2 a_2^2 e_2}
\frac{\partial R_2}{\partial e_2}
\right) \nonumber\\
&\quad - \frac{\sqrt{1-e_2^2}}{n_2 a_2^2 e_2}
\frac{\partial R_2}{\partial e_2}.
\end{align}
This equation, along with Equations \eqref{eq:dadt} and \eqref{eq:dedt}, completes the set of equations for $(a_2,e_2,\phi_2)$.

The conservative dynamics is conveniently understood using the Hamiltonian formalism. We can perform a canonical change of variables from the original Poincare set $(\lambda_2,\Lambda_2,\gamma_2,\Gamma_2)=(M_2 + \varpi_2,\sqrt{Gm_0(1-\beta)a_2}, -\varpi_2,\sqrt{Gm_0(1-\beta)a_2}(1-\sqrt{1-e_2^2})$ to $(\phi_2,\Gamma_2,Q,J_2)$ where $J_2=\sqrt{Gm_0(1-\beta)a_2}(j\sqrt{1-e_2^2}-j+k)$ \citep{2002clme.book.....G}. Here $M_2$ is the dust particle mean anomaly. The pairs $(\phi_2,\Gamma_2)$ and $(Q,J_2)$ form canonical angle–action pairs. The Kamiltonian of the new variable set is
\begin{eqnarray}\label{eq:Kamiltonian}
K & = &  -\frac{Gm_0(1-\beta)}{2a_2} - n_1\sqrt{Gm_0(1-\beta)a_2(1-e_2^2)} - R_2 
\nonumber \\ & \equiv &  - \frac{Gm_0}{2a_1} C_J,
\end{eqnarray}
where the dimensionless version $C_J$ will be used for convenience. The quantity $K$ is the Jacobi constant, here expressed in terms of the orbital elements. If the $Q$-averaged $\langle R_2 \rangle$ is used, and a circular inner orbit assumed, then $J_2$ is a constant of motion, as $Q$ no longer appears in $K$. It is convenient to work with a dimensionless form of $J_2 \equiv k\sqrt{Gm_0(1-\beta)a_{\rm 2, res}(1 + \kappa_2)}$, where
\begin{equation}\label{eq:kappa2}
\kappa_2 \equiv \frac{a_2}{a_{2,\mathrm{res}}}\left(\frac{j\sqrt{1-e_2^2}-(j-k)}{k}\right)^2-1.
\end{equation}
When PR drag is neglected, $\kappa_2$ is a constant of motion, implying that variations in $a_2$ and $e_2$ are correlated. Physically, $\kappa_2$ measures the proximity to exact resonance \citep{1999ssd..book.....M}. If $e_2 = 0$ and $a_2 = a_{2,\mathrm{res}}$, then $\kappa_2 = 0$. Far outside the resonance the dust orbit remains nearly circular with $a_2>a_{2,\mathrm{res}}$, giving $\kappa_2>0$. As the
particle migrates inward and enters resonance, $\kappa_2$ decreases and
becomes negative.
If we approximate $a_2 \simeq a_{\rm 2, res}$, then for $\kappa_2 < 0$ the eccentricity is approximately $e_2 \simeq \sqrt{k|\kappa_2|/j}$, showing the growth of $e_2$ with $|\kappa_2|$. 

When orbit averaged PR drag is included, $\kappa_2$ varies as
\begin{align}\label{eq:kappa2dot}
\dot{\kappa}_2
&= \left(1 + \kappa_2 \right)
   \left( \frac{\beta G m_0}{a_2^2 c} \right) \nonumber\\
&\quad \times \left[
-2 \left( \frac{1 + \tfrac{3}{2} e_2^2}{(1-e_2^2)^{3/2}} \right)
+ \frac{5 j e_2^2}{
\left[ j \sqrt{1-e_2^2} - j + k \right]
\left(1-e_2^2\right)}
\right].
\end{align}
During the evolution, $\kappa_2$ typically varies slowly compared to $\phi_2$ and $\Gamma_2$, and therefore labels a sequence for the evolution. In particular, libration around stable ``conservative equilibrium points" $(e_{2}(\kappa_2),\phi_{2}(\kappa_2))$, found by setting $\dot{a}_2=\dot{e}_2=0$ at fixed $\kappa_2$ and ignoring the PR terms, efficiently describes the evolution. If the conservative equilibrium points are sufficiently stable, the system will proceed to the ``dissipative equilibrium" point $(e_{2,\rm eq},\phi_{2,\rm eq},\kappa_{2,\rm eq})$ which is defined by setting the derivatives of $a_2$, $\phi_2$, and $\kappa_2$ to zero in Eqs.~\eqref{eq:dadt}, \eqref{eq:dphidt}, and \eqref{eq:kappa2dot}, including the PR terms. Setting Equation \eqref{eq:kappa2dot} to zero selects a particular value of $e_{2, \rm eq}$, which depends only on the resonance $j\!:\!j-k$ \citep{1994Icar..110..239B}. Given this value of $e_{\rm 2, eq}$, and assuming $a_2 \simeq a_{\rm 2, eq}$, the value of $\kappa_{2, \rm eq}$ can be found from Equation \eqref{eq:kappa2}. Writing $K=K_0 - R_2$, it can be shown that evolution under PR drag gives $\dot{K}_0=0$ at this same value of $e_{\rm 2, eq}$, so that both $\dot{\kappa}_2=0$ and $\dot{K}_0=0$ at the dissipative equilibrium point.

\subsection{Equilibrium Points}

\begin{figure*}[ht!]
    \centering
    \includegraphics[width=\textwidth]{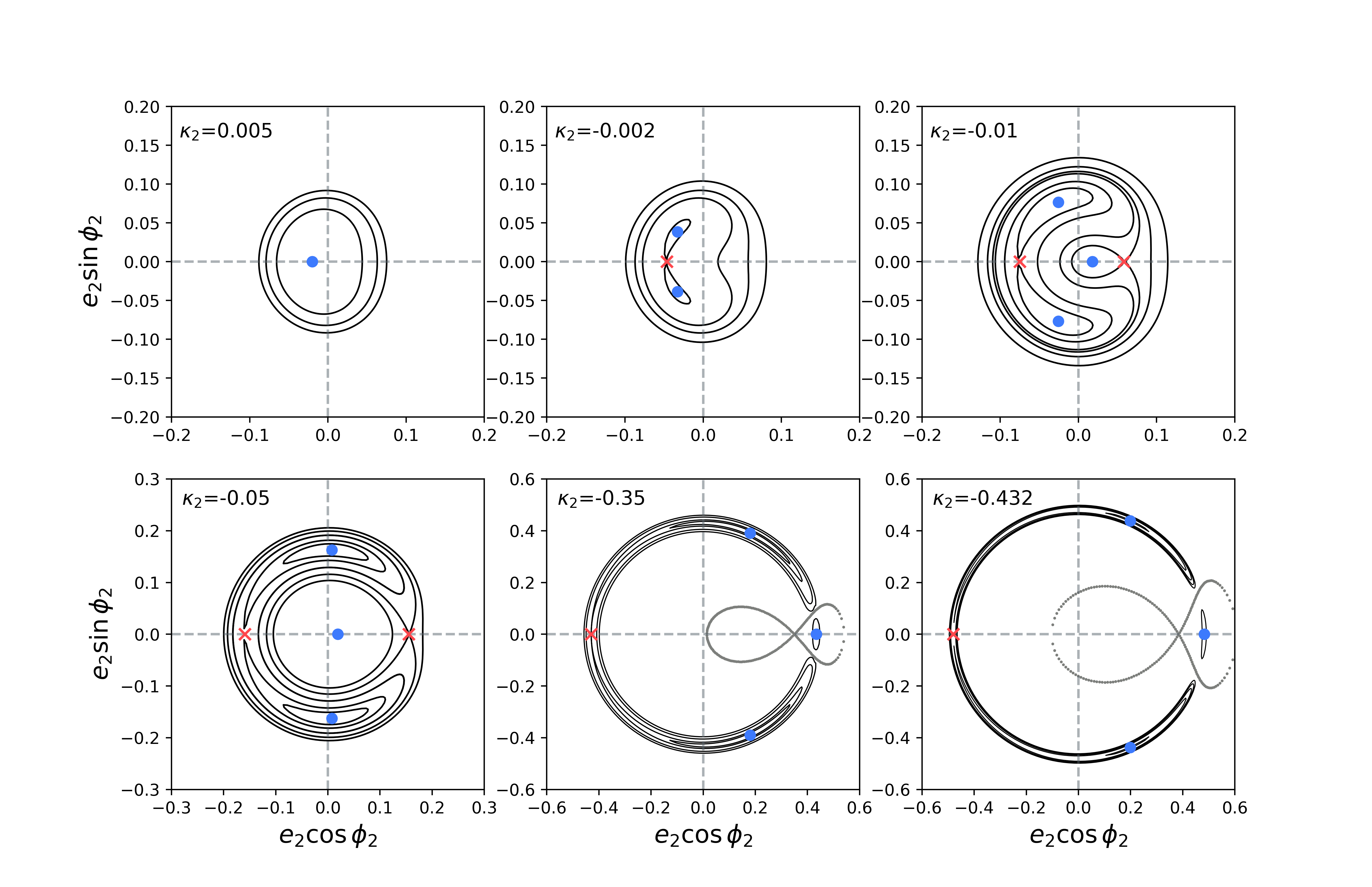}
    \caption{Contour plot of the Q-averaged Kamiltonian (Equations \ref{eq:Kamiltonian} and \ref{eq:Q_ave_R}) for stellar mass $m_0=1\, \rm M_\odot$ and planet mass $m_1 = 1\, M_J$. Semi-major axis $a_2=a_2(e_2,\kappa_2)$. Stable and unstable fixed points are marked by blue dots and red crosses, respectively. PR-driven evolution proceeds from left to right along the top and then bottom rows. Grey dots indicate the collision curve, corresponding to regions where the dust particle and planet orbits cross. The bottom right panel has the value of $\kappa_2 = -0.432$, corresponding to the dissipative equilibrium.
    \label{fig:phasespace1}}
\end{figure*}

\begin{figure*}[ht!]
    \centering
    \includegraphics[width=\textwidth]{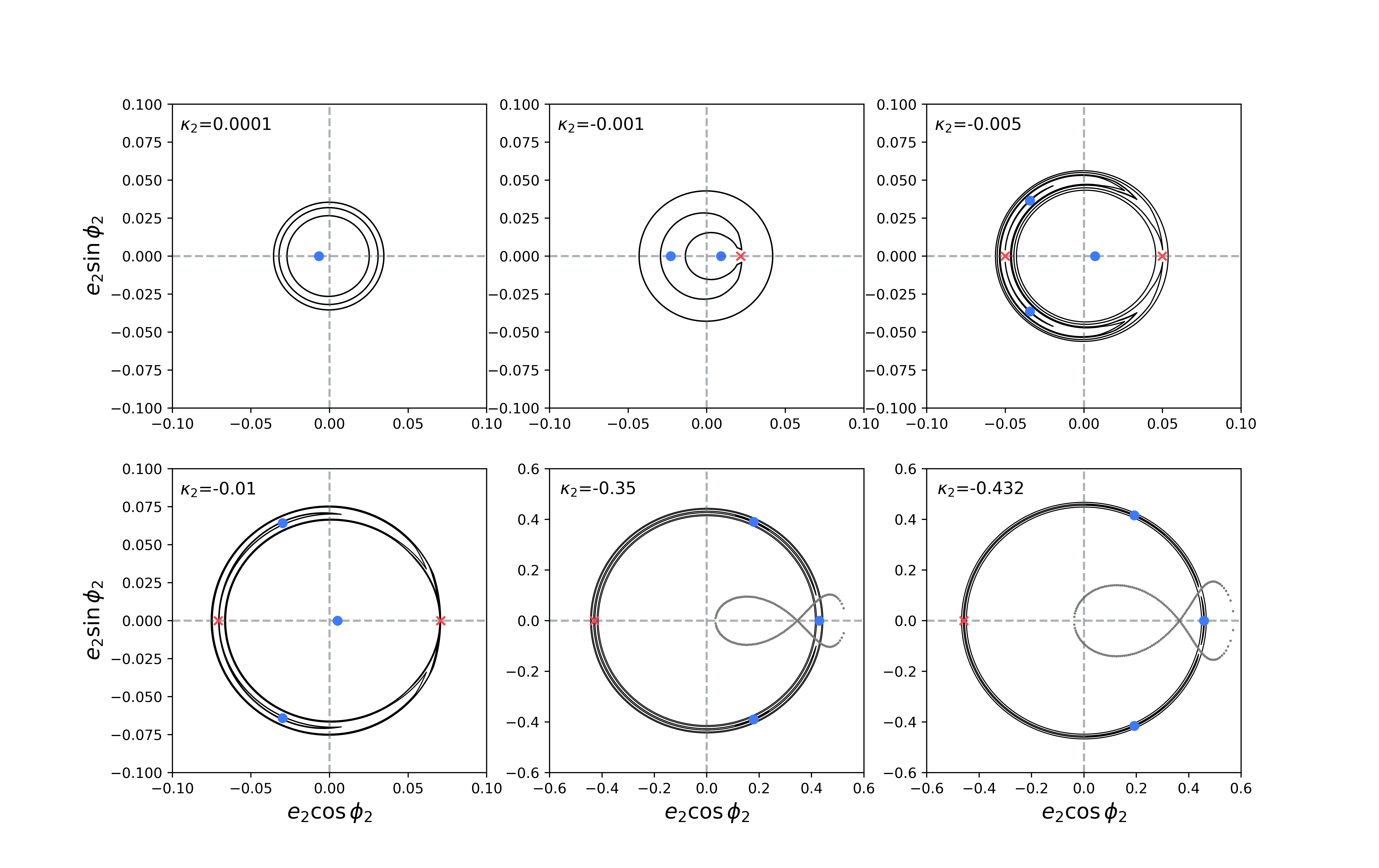}
    \caption{Same as Figure \ref{fig:phasespace1} 
    but with planet mass $m_1 = 10^{-2} M_J$.
    \label{fig:phasespace2}}
\end{figure*}

In this subsection, we analyze the phase-space structure and equilbrium points near the $2\!:\!1$ MMR. The level curves of the  averaged Kamiltonian $K$ are shown in Figure~\ref{fig:phasespace1}, for $m_0=1\, M_{\odot}$ and $m_1=1\, M_J$, and 
Figure \ref{fig:phasespace2}, for the same stellar mass but smaller planetary mass $m_1=10^{-2}\, M_J$. These two cases illustrate the difference in equilibrium points for planet masses above and below the critical value $m_1/m_0 \sim 10^{-4}$. 

Each panel in Figure~\ref{fig:phasespace1} shows the level curves of $\langle K\rangle (e_2,\phi_2,\kappa_2)$ for a different value of $\kappa_2$, with PR-driven evolution proceeding from left to right in the top and then bottom rows. For sufficiently slow migration rate, the trajectory will follow the conservative equilibrium points (blue points). 
Initially (top left), the only equilibrium point is on the -x axis ($\phi_2=\pi$), with $e_2$ increasing as $\kappa_2$ decreases. In this high planet mass case, this initially stable ``on-axis" equilibrium point becomes unstable, and two new stable ``off-axis" equilibrium points appear (top middle) at $\kappa_{2, \mathrm{off-axis}} \simeq 4.68\times 10^{-4}$ (see Appendix \ref{app:loworderexpansion} for the derivation of the bifurcation point). Thereafter, the evolution switches from on- to off-axis, following the stable equilibrium point, which in simulations we find is usually in the lower-half plane. 

As $\kappa_2$ decreases further, these off-axis points rotate around from the second and third quadrants to the first and fourth, respectively. New equilibrium points appear on the +x axis at $\kappa_{2, \mathrm{on-axis}} \simeq -6.57\times 10^{-3}$, with one moving to smaller $e_2$ and one to larger $e_2$. The small (large) $e_2$ solution is initially stable (unstable). By this point, there are four solutions with large and nearly equal $e_2$, and one solution near the origin with small $e_2$ (top right and bottom left). 

When the eccentricity becomes sufficiently large, the dust and planet orbits cross, at which point $R_2$ may diverge. The dotted ``collision curve" is the locus of points where $\vb*{x}_1=\vb*{x}_2$, and is found in the high $e_2$ cases in the bottom center and bottom right panels. Note that the collision curves do not strictly correspond to guaranteed collisions, since the Kamiltonian has been averaged over fast angles. Rather, they indicate regions where the orbit approaches the planet closely and where the averaged description may break down due to close encounters. Our implementation of the collision curve differs from that of \citet{1994Icar..110..239B}, where they use $a_2 \simeq a_{2, \rm res}$. In contrast, we use $a_2 = a_2(\kappa_2,e_2)$, so it varies along phase-space trajectories and the collision condition must be evaluated self-consistently rather than imposed at a fixed resonant $a_2$. 

Once the collision curve appears, the initially unstable equilibrium point on the $+x$ axis becomes stable. However, none of the simulated trajectories follow this branch, likely because reaching it requires crossing the collision curve. In general, we may expect that close encounters occur for large libration amplitude around the two off-axis stable equilibrium points, as they approach the collision curves. During such close encounters, the $Q$-averaged $\langle R_2 \rangle$ is insufficient as it does not include the terms in $R_2$ needed at close separation. Nevertheless, the dust particle escapes resonance by crossing the separatrix as a result of PR drag. In some cases, close encounters may assist the escape process, but PR drag remains the underlying mechanism that drives the particle out of resonance; close encounters alone do not initiate the escape. A detailed explanation, together with illustrative examples, is presented in subsection~\ref{sec3.2}.

Figure \ref{fig:phasespace2} shows the level curves for the small planet case with the same  $m_0=1\, M_{\odot}$ but $m_1 = 10^{-2} M_J$. While the phases of just entering resonance (top left) and high $e_2$ evolution (large negative $\kappa_2$, bottom row)  look the same, the equilibrium points at small $e_2$ differ. In this case, the on-axis bifurcation occurs first (top middle) at $\kappa_{2, \mathrm{on-axis}} \simeq -2.95 \times 10^{-4}$, creating a stable (small $e_2$) and unstable (large $e_2$) pair of equilibrium points on the +x axis. In simulations we find that the evolution continues to follow the stable equilibrium point on the -x axis. Next, that -x equilibrium point becomes unstable at $\kappa_{2, \mathrm{off-axis}} \simeq -1.78 \times 10^{-3}$ and two stable off-axis equilibrium points are created at the same time. The later evolution follows these off-axis equilibrium points, and subsequent evolution resembles the high mass case. The closed libration curves surrounding the stable equilibrium points appear smaller in this case, as their width scales with the planet mass ($m_1$).

When PR drag is included, the evolution will only follow the stable equilibrium points closely when the migration rate $\tau_{\rm n2}^{-1}$ is much smaller than the libration frequency around the fixed point (see Appendix \ref{app:loworderexpansion}). This clearly breaks down when the evolution switches from the on- to off-axis equilibrium points, as the libration frequency goes to zero at that point. We will show that the evolution can drift away from the equilibrium points until the libration frequency has increased sufficiently, which occurs when the points have separated away from the -x axis sufficiently.

\section{Simulation Results}\label{sec3}

The dynamical evolution can be summarized as follows. Initially, PR drag induces a gradual loss of energy and angular momentum from the orbit, causing the dust to slowly spiral inward on nearly circular orbits. In the process of orbital decay, for sufficiently slow migration rate, the dust particle can be trapped in mean motion resonances with the planet. However, with dissipative effects driven by PR drag, dust particles will ultimately escape resonance trapping because of overstable libration. 

In this section, we present numerical integrations of the EEOM to illustrate the time evolution of the orbital elements from the initial orbital decay, to capture into resonance, and then escape from resonance.
We include comparisons to numerical solution of Lagrange's Planetary Equations, with different levels of approximation for $R_2$ to understand how high $e_2$ affects the evolution in Appendices \ref{app:firstorder} and \ref{app:secondorder}.

\subsection{Capture into $2\!:\!1$ MMR }

\begin{figure*}[ht!]
    \centering
    \includegraphics[width=\textwidth]{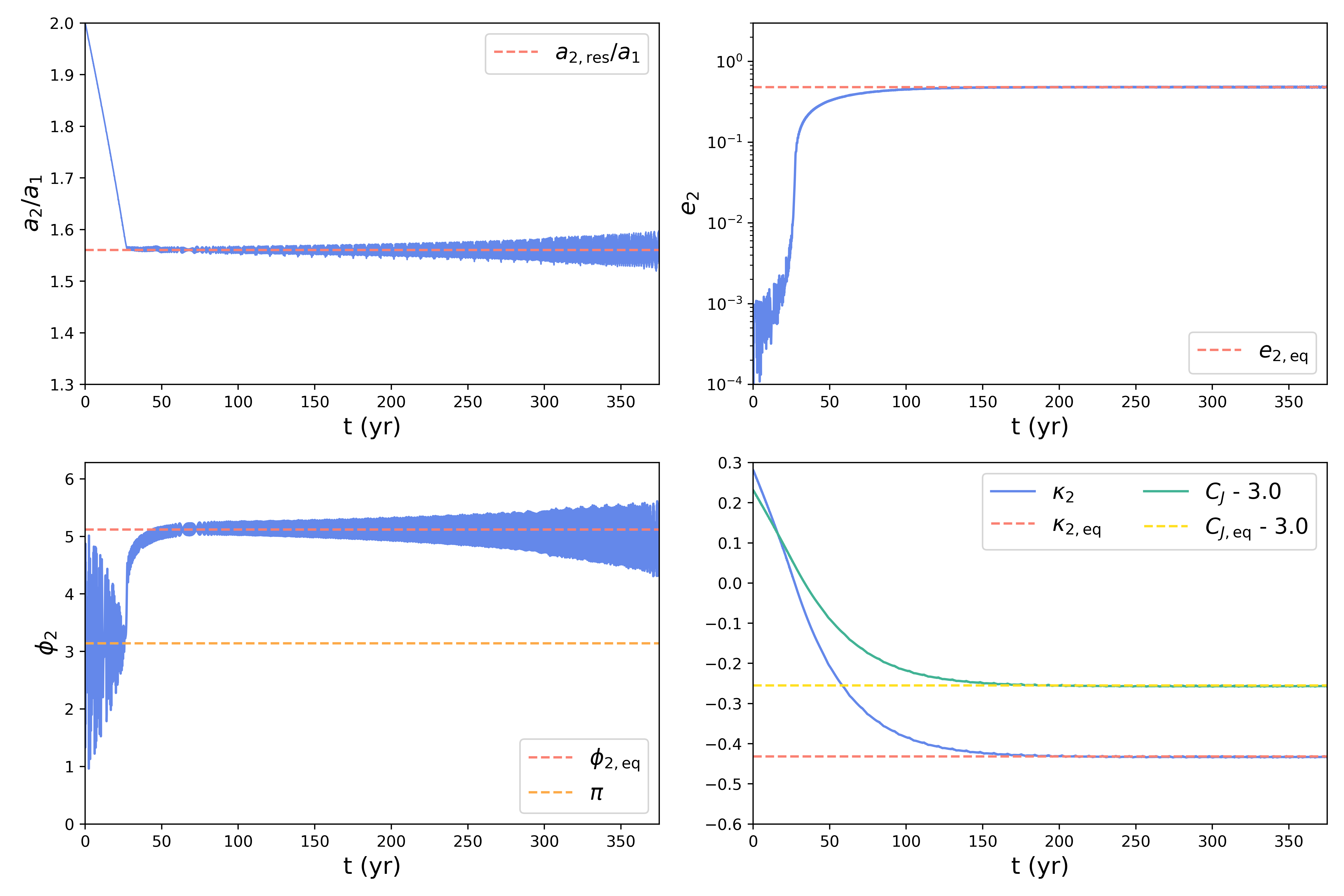}
    \caption{EEOM numerical solution showing the $2\!:\!1$  MMR. Parameters used are $m_0=1\, M_{\odot}$, $m_1 = 1\, M_J$, $a_1=0.05 \,\mathrm{AU}$ and $\beta = 0.05$. The upper left, upper right, lower left and lower right panels show the semi-major axis $a_2$, eccentricity $e_2$, resonant angle $\phi_2$, Jacobi constant $C_J \propto K_2$, and resonance constant of the motion $\kappa_2$ as a function of time. Dashed lines are values for the conservative or dissipative equilibrium points. 
    \label{fig:reboundplot}}
\end{figure*}

Figure~\ref{fig:reboundplot} illustrates a typical trajectory showing evolution through the $2\!:\!1$ MMR. We consider a system consisting of a Sun-like star ($m_0 = 1\, M_{\odot}$) and a Jupiter-like planet ($m_1 = 1\, M_J$) orbiting at $a_1 = 0.05\,\mathrm{AU}$. The particle is initialized on a circular orbit well outside resonance at $a_2 = 5 a_1$. The upper-left panel shows the semi-major axis evolution of the dust particle. Initially it undergoes orbital decay due to PR drag. After entering resonance, it oscillates around $a_{2,\mathrm{res}}$
with increasing amplitude until it eventually escapes. The upper-right panel shows the corresponding evolution of eccentricity, where $e_2$ increases from an initially small value up to $e_{\rm 2} \sim 0.48$, after which it oscillates with increasing amplitude. 

The evolution proceeds in two distinct stages. In the first stage, the particle follows the $\phi_2=\pi$ conservative fixed point seen in Figure~\ref{fig:phasespace1}. As $\kappa_2$ decreases, the eccentricity increases gradually to $e_2 \simeq 0.03$, while the resonant angle $\phi_2$ librates about $\pi$ (lower-left panel). In the second stage, the particle switches to the off-axis fixed point, with $e_2$ approaching its dissipative equilibrium value $e_{2,\mathrm{eq}} \simeq 0.4812$ and $\phi_{2,\mathrm{eq}} \simeq 5.1464$
as also seen in Figure~\ref{fig:phasespace1}. Both $\kappa_2$ and $C_J$ evolve slowly in time, labeling the sequence of conservative equilibrium points, until the dissipative equilibrium is reached; thereafter they remain more nearly constant in time (lower-right panel).

\begin{figure}[ht!]
    \centering
    \includegraphics[width=\columnwidth]{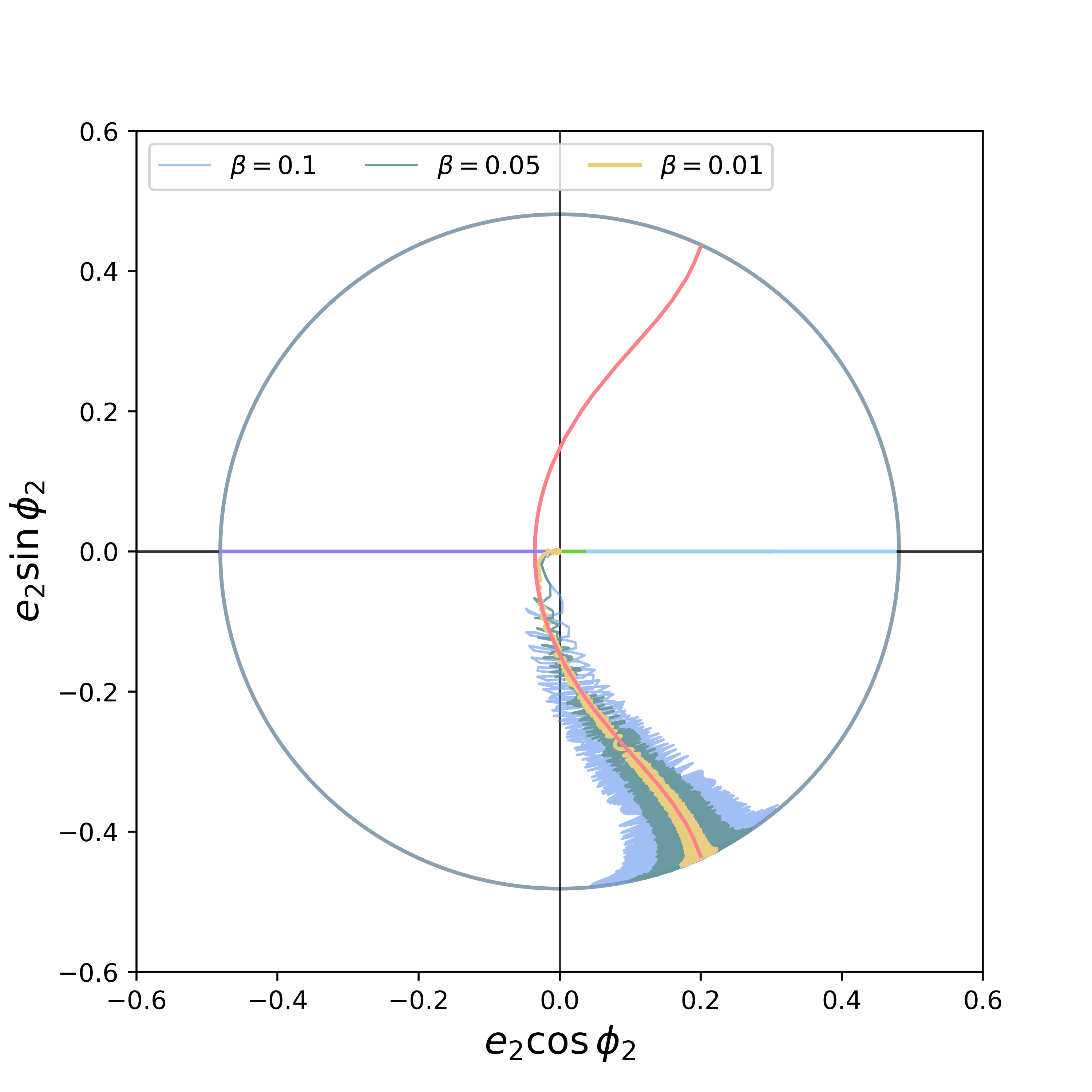}
    \caption{Dust particle trajectories in phase space for different values of $\beta$ with $m_0 = 1\, M_{\odot}$, $m_1 = 1\, M_J$, and $a_1=0.05 \rm AU$. Trajectories with $\beta=0.1$, $0.05$, and $0.01$ are shown in blue, green, and yellow, respectively. The locations of the fixed points are indicated by colored curves: red (two off-axis points), purple ($-x$ axis), green ($+x$ axis near the origin), and blue ($+x$ axis farther from the origin). The gray circle marks the dissipative equilibrium at $e_{2,\rm eq} \simeq 0.4812$.}
    \label{fig:reboundplotphasespace1}
\end{figure}

\begin{figure*}[ht!]
    \centering
    \includegraphics[width=\textwidth]{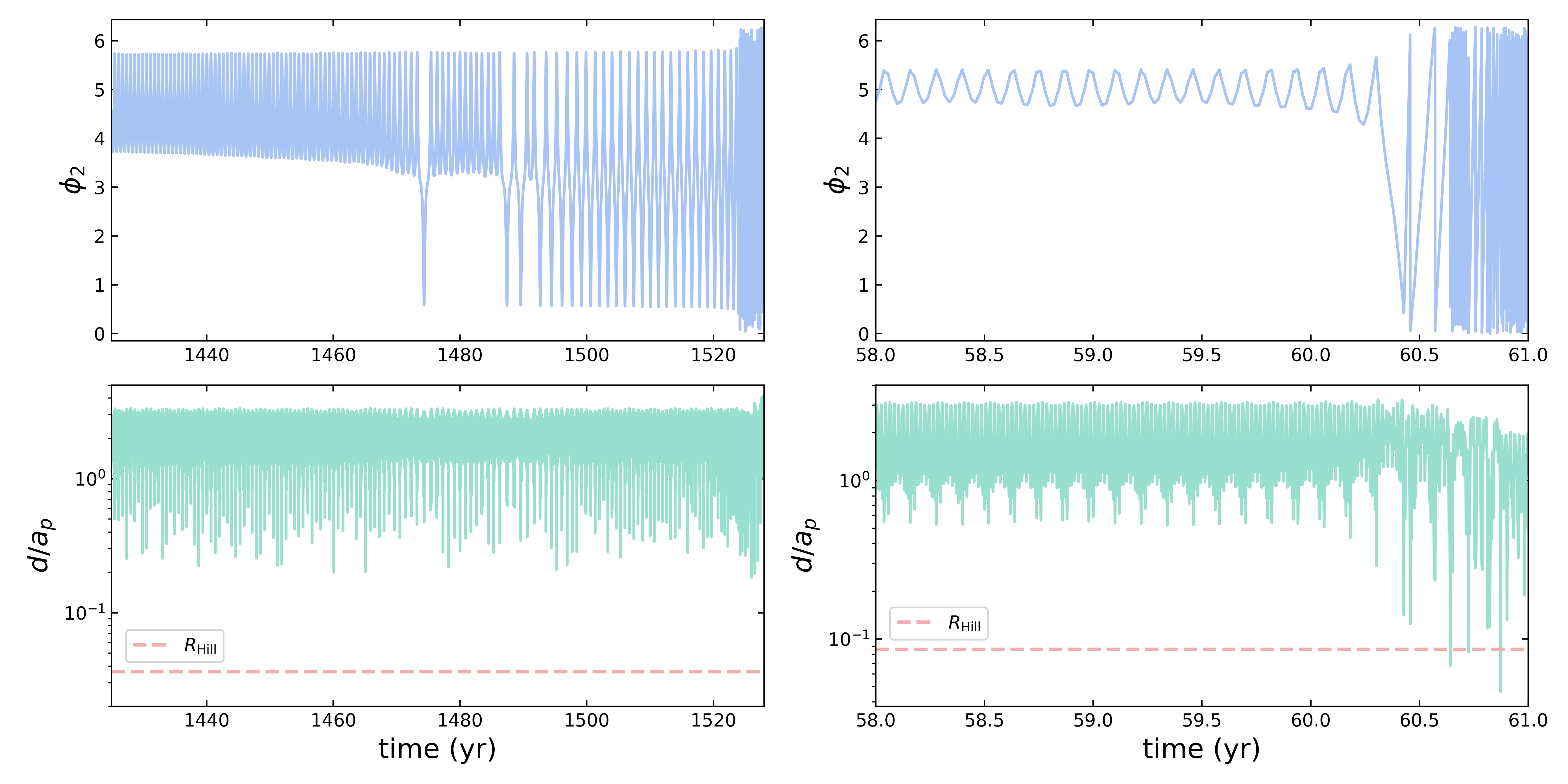}
    \caption{Examples of smooth (left) and encounter-assisted (right) escape from the $2\!:\!1$ MMR, shown through the evolution of the resonant angle $\phi_2$ (top) and the planet–dust separation $d$ (bottom). The left panels correspond to $m_1 = 0.15\,M_J$ and $\beta = 0.01$, while the right panels correspond to $m_1 = 2\,M_J$ and $\beta = 0.25$. In both cases, we adopt $m_0 = 1\, M_\odot$ and $a_1 \simeq 0.05\,\mathrm{AU}$. The dashed line indicates the Hill radius.
    \label{fig:escape_resonance}}
\end{figure*}

Three trajectories for $\beta=0.01$, $0.05$, and $0.1$ with $m_0 = 1\, M_{\odot}$ and $m_1 = 1\, M_J$
are shown in Figure \ref{fig:reboundplotphasespace1}. 
The  particle initially follows a fixed point along the $-x$ axis, representing the first stage of resonance capture. 
When an off-axis bifurcation occurs and the original fixed point becomes unstable, two new stable off-axis fixed points emerge. However, the particle does not immediately follow these new points. Instead, it gradually departs from the $-x$ axis due to the libration timescale being much longer than the migration timescale.
As the libration timescale decreases, the dust particle eventually catches up and begins to follow one of the off-axis fixed points, librating around it until reaching dissipative equilibrium. 
The faster the migration rate (i.e., the larger $\beta$), the longer it takes for the particle to catch up with the evolving off-axis fixed points. A detailed analysis of the stability of these fixed points is presented in Section~\ref{sec4}, using both analytical and numerical methods, and demonstrates that dust particles ultimately escape from resonance.

\subsection{Escape from Resonance}
\label{sec3.2}

Resonant capture under PR drag is temporary: dissipative forces drive overstable librations whose amplitude grows until the particle leaves the resonance. \citet{2015MNRAS.448..684S} suggested that dust–planet close encounters are responsible for resonance escape, with particles leaving resonance at a fixed encounter distance. However, as shown by our fixed point stability analysis in Section~\ref{sec4}, resonance trapping is intrinsically unstable even when only the Q-averaged $\langle R_2 \rangle$ 
is retained (see Equation~\ref{eq:growthrate_analytic}), and thus does not require close encounters. Dust particles therefore escape from resonance generically; close encounters may accelerate this process but are not required for escape to occur. Our numerical integrations reveal two distinct escape channels, depending primarily on the planet mass $m_1$ and the radiation pressure parameter $\beta$. 

Representative examples of these outcomes are shown in Figure~\ref{fig:escape_resonance}.
For low planet mass and small $\beta$, escape is ``smooth'': the  libration amplitude increases gradually and the particle crosses the separatrix without approaching the planet. In this regime, the planet–particle separation remains large throughout the escape, with $d \gg r_{\rm Hill}$, where $r_{\rm Hill} = a_1 (m_1 / (3m_0))^{1/3}$ is the Hill radius. The left column of Figure~\ref{fig:escape_resonance} illustrates such a case, with $m_1 = 0.15\,M_J$ and $\beta = 0.01$, using $m_0 = 1 M_\odot$ and $a_1 \simeq 0.05\,\mathrm{AU}$. As the particle escapes resonance, it first transitions between two off-axis fixed points (Figure \ref{fig:phasespace1}) while continuing to librate, before the resonant angle ultimately begins to circulate. Throughout this process, the planet–particle distance remains well outside the Hill radius.

For larger planet mass $m_1$, resonance escape can be assisted by a close encounter with the planet. In these cases, the planet-dust separation $d$ decreases to a few $r_{\rm Hill}$, coincident with an abrupt change in the orbital elements and the termination of libration. The right column of Figure~\ref{fig:escape_resonance} shows an example with $m_1 = 2\,M_J$ and $\beta = 0.25$, using the same stellar mass and semi-major axis. Here, the resonant angle ceases libration at the time of the close encounter (when the dust particle passes within a few Hill radii of the planet) and subsequently begins rapid circulation, marking escape from resonance.

The physical mechanism by which close encounters change the orbit from libration to circulation remains to be understood. If $C_J$ is constant during the encounter, one possibility is that $\kappa_2$ is sufficiently changed, as it is not an exact constant of the motion for the exact $R_2$, such that the level curve of $C_J$ at the new $\kappa_2$ is has switched from libration to circulation. We leave this for a future investigation.

\begin{figure}[!b]
    \centering
    \includegraphics[width=\columnwidth]{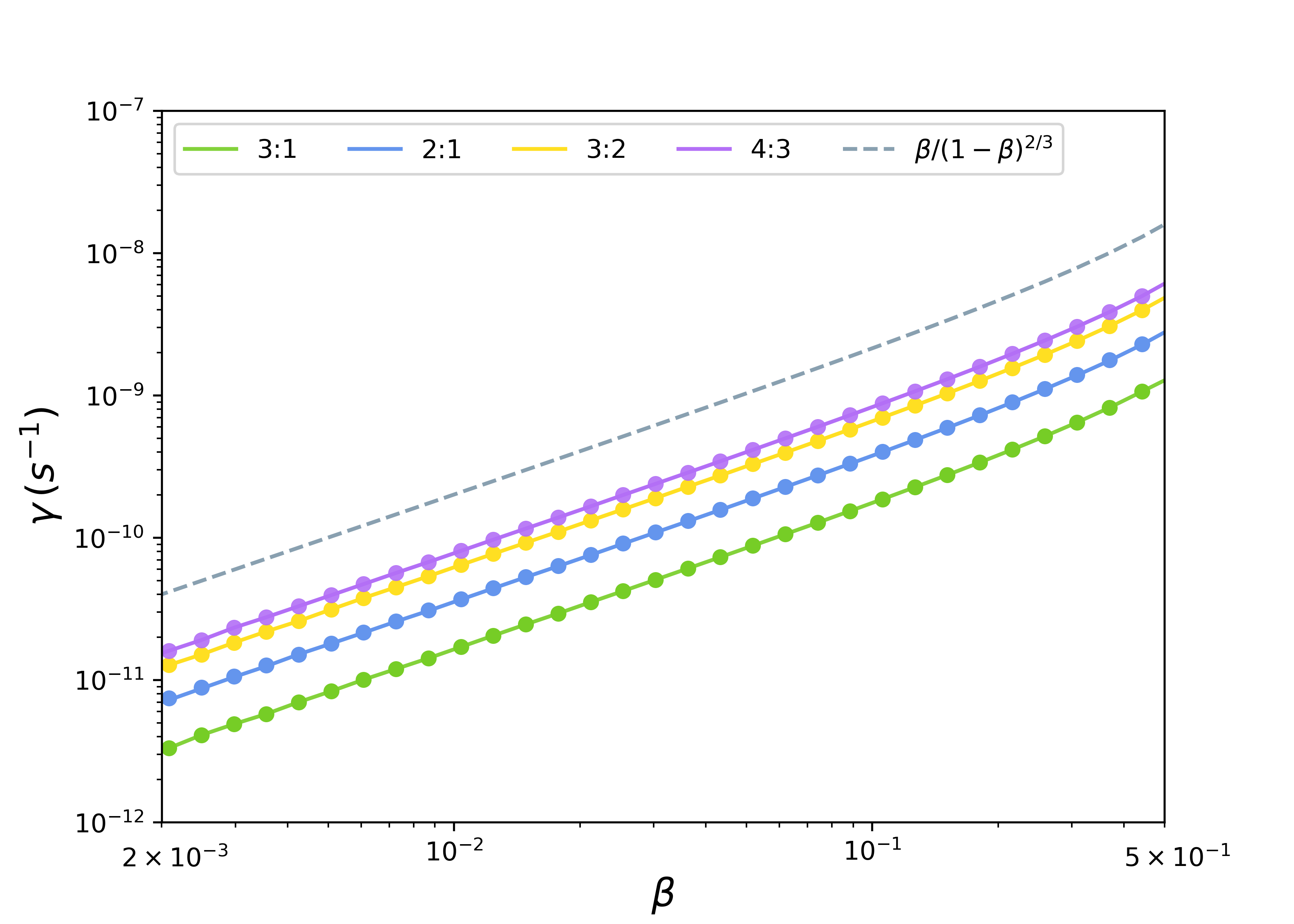}
    \caption{Growth rates $\gamma>0$ for small perturbations around dissipative equilibrium points for different MMRs as a function of $\beta$. All growth rates are positive, indicating overstability.
    Solid lines show numerical linearization results (Equation \ref{eq:Aij_numerical}); dotted lines show semi-analytical solutions (Equation \ref{eq:growthrate_analytic}). The parameters used in this figure are $m_0 = 1\, M_{\odot}$, $m_1 = 1\, M_J$, and $a_1=0.05 \rm AU$.
    }
    \label{fig:growthrate}
\end{figure}

\section{Stability of Dissipative Equilibrium Points}\label{sec4}

In Appendices \ref{app:firstorder} and \ref{app:secondorder}, we review the ${\cal O}(e_2)$ and hybrid-${\cal O}(e_2^2)$ approximations to LPE 
for $n_2$, $e_2$ and $\phi_2$. The conservative and dissipative equilibrium points are found,
and the equations are linearized to study the libration frequency and growth rate of small displacements due to PR drag.
However, the results for the dissipative equilibrium points 
using ${\cal O}(e_2)$ or ${\cal O}(e_2^2)$ LPE found them to be stable,
which contradicts the exact EEOM results where the particle eventually escapes from resonance. This is an important issue, as if particles could build up at MMRs we would expect to find large dust densities there. Further experimentation found that including harmonics up to ${\cal O}(e_2^4)$ is necessary to find the overstability. Inclusion of so many terms in an analytic treatment is unwieldy as it requires computation of many Laplace coefficients, and a more compact treatment is desired. 

In this section, we use the $Q$-averaged $\langle R_2 \rangle $ from Equation \eqref{eq:Q_ave_R} and the exact LPE and PR drag terms to numerically solve for the dissipative equilibrium points. We then use both numerical and analytic approaches to study the libration frequency and growth rate of small displacements at these points due to PR drag. 

We begin with the numerical approach. The dissipative equilibrium points are obtained by setting $\dot{n}_2=0$, $\dot{e}_2=0$, and $\dot{\phi}_2=0$ using Equations \eqref{eq:dadt}, \eqref{eq:dedt}, and \eqref{eq:dphidt}, where $\dot{n}_2 = -3n_2 \dot{a}_2/(2a_2)$. For each resonance and value of $\beta$, we use a multidimensional root-finding algorithm initialized near $n_{2, \rm guess}=(j-k)n_1/j$, $e_{2, \rm guess}$ obtained numerically from $\dot{\kappa}_2=0$ in Equation \eqref{eq:kappa2dot}, and $\phi_{2, \rm guess}=3\pi/2$ to solve for the equilibrium values $n_{2, \rm eq}$, $e_{2, \rm eq}$, and $\phi_{2, \rm eq}$. To study the local stability of these equilibrium points, we numerically compute the Jacobian matrix by finite-differencing the exact equations of motion about each equilibrium point:
\begin{equation} \label{eq:Aij_numerical}
A_{ij}=\frac{\partial \dot x_i}{\partial x_j},
\qquad
{\bf x}=(n_2,e_2,\phi_2).
\end{equation}
The eigenvalues $s=-i \omega + \gamma$ of this matrix determine the libration frequency $\omega$ and growth rate $\gamma$
of small displacements from the equilibrium point. We use \texttt{SciPy} \citep{2020SciPy-NMeth} to find the equilibrium points and to compute the eigenvalues of the finite-difference Jacobian. In Figure~\ref{fig:growthrate}, we plot the numerically computed $\gamma$ as a function of $\beta$, shown as solid curves with different colors corresponding to different MMRs. Throughout this calculation, we adopt $m_0 = 1\, M_{\odot}$, $m_1 = 1\, M_J$, and $a_1=0.05 \rm AU$. For all resonances and values of $\beta$ shown, the growth rates are positive, indicating that the dissipative equilibrium points are locally unstable. This is consistent with the instability of dissipative equilibrium found by \citet{1994Icar..110..239B}.

To understand these numerical results for the growth rate, we now turn to an analytic approach. We replace $n_2$ with $\kappa_2$, since variations in $n_2$ and $e_2$ are related, whereas $\kappa_2$ evolves only due to PR drag effects when using $\langle R_2 \rangle$. We further simplify the derivation by switching from $e_2$ to the action $\Gamma_2$ and writing the Hamilton's equations of motion for the action-angle variables. 

\begin{table*}[htbp]
\centering
\caption{Dissipative equilibrium values for $e_{2,\rm eq}$, $\phi_{2,\rm eq}$, and $\kappa_{2,\rm eq}$, libration frequency $\omega_0/n_{2,\rm eq}$, and growth rate coefficients from Equation \ref{eq:gamma_formula}
for the oscillatory ($\gamma_{j:(j-k)}$) and non-oscillatory ($\gamma_{\rm non\text{-}osc,\,j:(j-k)}$) modes.
The integers $j$ and $k$ label the resonance.
The parameters used in this table are $m_0 = 1\, M_{\odot}$, $m_1 = 1\, M_J$, and $a_1=0.05 \rm AU$.}
\begin{tabular}{ccccccc}
\hline
$j:j-k$ & $e_{2,\rm eq}$ & $\phi_{2,\rm eq}$ & $\kappa_{2,\rm eq}$ & $\omega_0/n_{2, \rm eq}$ & $\gamma_{j:(j-k)}$ & $\gamma_{\rm non\text{-}osc,\,j:(j-k)}$\\
\hline
$3\!:\!1$ & $0.5993$ & $0.58,\ 2.56$ & $-0.5088$ & $0.14$ & $0.179$ & $-1.22$\\
$2\!:\!1$ & $0.4812$ & $1.14,\ 5.14$ & $-0.4326$ & $0.11$ & $0.385$ & $-2.12$\\
$3\!:\!2$ & $0.3690$ & $\pi$ & $-0.3786$ & $0.052$ & $0.672$ & $-3.22$\\
$4\!:\!3$ & $0.3108$ & $\pi$ & $-0.3569$ & $0.074$ & $0.845$ & $-3.83$\\
\hline
\end{tabular}
\label{tab:mmr_eq}
\end{table*}

Using the Kamiltonian $K = K(\Gamma_2, \phi_2, \kappa_2)$ from Equation \eqref{eq:Kamiltonian}, the equations of motion including PR drag terms are
\begin{equation}\label{eq:dgamma2dt_full}
\frac{d\Gamma_2}{dt} = -K_{,\phi} + \dot{\Gamma}_2,
\end{equation}
\begin{equation}\label{eq:dphi2dt_full}
\frac{d\phi_2}{dt} = K_{,\Gamma},
\end{equation}
\begin{equation}\label{eq:dkappa22dt_full}
\frac{d\kappa_2}{dt} = \dot{\kappa}_2.
\end{equation}
Here the comma denotes a partial derivative, $\dot{\kappa}_2=\dot{\kappa}_2(\Gamma_2,\kappa_2)$ due to PR drag is given in Equation \eqref{eq:kappa2dot}, and a similar formula for $\dot{\Gamma}_2 = \dot{\Gamma}_2(\Gamma_2, \kappa_2)$ is obtained using the PR drag terms in Equations \eqref{eq:dadt} and \eqref{eq:dedt}.

The dissipative (conservative) equilibrium points are found by setting the time derivative terms to zero, including (neglecting) the dissipative terms. As we now use the numerically $Q$-averaged $\langle R_2 \rangle$, we perform numerical root solving to find these equilibrium points using \texttt{SciPy} \citep{2020SciPy-NMeth}. The conservative equilibrium values of $e_{2,\rm eq}$, $\phi_{2,\rm eq}$, and $\kappa_{2,\rm eq}$ for the resonances shown in Figure~\ref{fig:growthrate} are summarized in Table~\ref{tab:mmr_eq}. For $n_2 t_{\rm pr} \gg 1$, the dissipative equilibrium points should be only slightly shifted from conservative equilibrium points. Here $t_{\rm pr}=a_2^2c/(\beta Gm_0)$ is an orbital decay timescale due to PR drag.

Given the dissipative equilibrium point, Equations \eqref{eq:dgamma2dt_full}--\eqref{eq:dkappa22dt_full} can be linearized to study the evolution of small displacements $(\delta \Gamma_2, \delta \phi_2, \delta \kappa_2)$, yielding the following linear system
\begin{equation}
\begin{pmatrix}
\delta \dot{\Gamma}_2 \\
\delta \dot{\phi}_2 \\
\delta \dot{\kappa}_2
\end{pmatrix}
=
\begin{pmatrix}
-K_{,\phi\Gamma} + \dot{\Gamma}_{,\Gamma} & -K_{,\phi\phi} & -K_{,\phi\kappa}+\dot{\Gamma}_{,\kappa}\\
K_{,\Gamma\Gamma} & K_{,\Gamma\phi} & K_{,\Gamma\kappa}\\
\dot{\kappa}_{,\Gamma} & 0 & \dot{\kappa}_{,\kappa}
\end{pmatrix}
\begin{pmatrix}
\delta \Gamma_2 \\
\delta \phi_2 \\
\delta \kappa_2
\end{pmatrix}
.
\end{equation}
Plugging in a time-dependence $e^{st}$, with $s=-i \omega + \gamma$, this eigenvalue problem takes the form $sx = (A_0+A_1)x$, where $x=(\delta \Gamma_2,\delta \phi_2,\delta \kappa_2)$ is the solution vector. Here $A=A_0+A_1$ is the total matrix: $A_0$ is the piece of the matrix ignoring PR drag terms, while $A_1$ is the piece solely due to PR drag. 

While the libration frequency and growth rate may be straightforwardly derived from the characteristic polynomial of the matrix $A$, it is instructive to calculate the eigenvectors, treating $A_1$ as a small perturbation. 
A complication is that $A_0$ is non-symmetric and the right ($s r=A_0 r$) and left ($s \ell =A_0^T \ell$) eigenvectors are not identical. They can nevertheless be organized into orthogonal pairs of left and right vectors.

In the absence of dissipation, the libration frequency is $\omega^2_0 = K_{,\Gamma\Gamma}K_{,\phi\phi}-{K^2}_{,\Gamma\phi}$. For stable solutions, this quantity is positive and curves of constant $K$ are ellipses around the equilibrium point. For unstable modes, $\omega_0^2<0$ and the corresponding curves are hyperbolas. These shapes of the level curves are shown in Figures \ref{fig:phasespace1} and \ref{fig:phasespace2}. The conservative matrix $A_0$ has two stable eigenmodes of $A_0$ with $\omega=\pm \omega_0$, and a third, non-oscillatory mode with $\omega=0$. For the mode with $\omega=\omega_0$, the (unnormalized) right and left eigenvectors are
\begin{eqnarray}
r & = & \left( 
\begin{array}{c}
K_{,\phi\phi} \\
i \omega_0 - K_{,\Gamma \phi} \\
0
\end{array}
\right)
\end{eqnarray}
and
\begin{eqnarray}
l & = & \left( 
\begin{array}{c}
- K_{,\Gamma\Gamma} \\
i \omega_0 - K_{,\Gamma \phi} \\
(i/\omega_0)\left[ K_{,\phi \kappa} K_{,\Gamma \Gamma} + K_{,\Gamma \kappa} \left( i \omega_0 - K_{,\Gamma \phi}\right) \right]
\end{array}
\right).
\end{eqnarray}
Notice that the right eigenvector has $\delta \kappa_2=0$,
as expected for libration  around a conservative equilibrium point, for which $\kappa_2$ is constant. By contrast, the right eigenvector of the $\omega=0$ modes has a component along $\delta \kappa$, while the corresponding left eigenvector is $l=(0,0,1)$. Thus the $\omega=0$ mode corresponds to changes in $\kappa_2$, which has no restoring force and hence have zero frequency.

The perturbation to the eigenvalue, denoted here by $s_1$, has the usual form of the matrix element of $A_1$, but here evaluated between the left and right eigenvectors:
\begin{eqnarray}
s_1 & = & \frac{l^T A_1 r}{l^T r}.
\end{eqnarray}
The denominator allows for non-normalized eigenvectors. The imaginary part of $s_1$ gives a correction to $\omega$ due to PR drag. Here we focus on the real part of $s_1$, which gives the leading-order contribution to the growth rate. For the mode with $\omega=\omega_0$, we find
\begin{eqnarray} \label{eq:growthrate_analytic}
\gamma & \simeq & \frac{1}{2}\dot{\Gamma}_{,\Gamma} + \frac{1}{2}\dot{\kappa}_{,\Gamma}
\left(\frac{K_{,\phi\phi}K_{,\Gamma\kappa} - K_{,\Gamma\phi}K_{,\phi\kappa}}{\omega_0^2}\right),
\end{eqnarray}
while for the mode with $\omega=0$ 
\begin{equation}
\gamma \simeq \dot{\kappa}_{,\kappa} - \dot{\kappa}_{,\Gamma} 
\left(\frac{K_{,\phi\phi}K_{,\Gamma\kappa} - K_{,\Gamma\phi}K_{,\phi\kappa}}{\omega_0^2}\right).
\end{equation}
These two expressions involve quantities that can be evaluated semi-analytically at the equilibrium point. However, for the large $e_2$ of interest, a significant simplification occurs.
We can write Equation \eqref{eq:Kamiltonian} as $K=K_0(\Gamma_2,\kappa_2) - R_2(\Gamma_2,\phi_2,\kappa_2)$, where $K_0$ is independent of $m_1$ while $R_2 \propto m_1$.The $R_2$ terms may be treated as small compared to $K_0$ terms when $e_2 \gg \mu_1^{1/3}$. In this limit, the ratio simplifies considerably to
\begin{eqnarray}
    \frac{K_{,\phi\phi}K_{,\Gamma\kappa} - K_{,\Gamma\phi}K_{,\phi\kappa}}{K_{,\Gamma\Gamma}K_{,\phi\phi}-{K^2}_{,\Gamma\phi}} & = & 
    \frac{R_{2,\phi\phi}K_{0,\Gamma\kappa} + R_{2,\Gamma\phi}R_{2,\phi\kappa}}{K_{0,\Gamma\Gamma}R_{2,\phi\phi}+{R^2}_{2,\Gamma\phi}}
    \nonumber \\ & \simeq & 
    \frac{K_{0,\Gamma\kappa} }{K_{0,\Gamma\Gamma}}.
\end{eqnarray}
Thus, to leading order in this regime, derivatives of $R_2$ are not needed in the approximate expression.

Given the numerical solutions for the dissipative equilibrium points using the $Q$-averaged $\langle R_2 \rangle$ and exact LPE, what remains is to take the analytic derivatives of $K_0$, $\dot{\Gamma}$, and $\dot{\kappa}$ for which we used \texttt{SymPy} \citep{Meurer2017SymPy}. The resulting growth rates can be written as
\begin{eqnarray} \label{eq:gamma_formula}
\gamma & \simeq & \gamma_{j:(j-k)} \left( \frac{Gm_0}{a_1^2c} \right) \left( \frac{\beta}{(1-\beta)^{2/3}} \right)
\end{eqnarray}
where all the dependence on parameters as been factored out, except for the dimensionless constant $\gamma_{j:(j-k)}$ which depends on the resonance. The dimensional prefactor is of order the PR orbital decay rate at the position of the planet. For the oscillatory mode with $\omega=\omega_0$ solution, we find overstability for all resonances considered. By contrast, the $\omega=0$ solution is always damped with $\gamma_{\rm non\text{-}osc,\,j:(j-k)} < 0$. 

The libration frequency $\omega_0$, the growth rate prefactor $\gamma_{j:(j-k)}$, and the prefactor of the non-oscillatory mode damping rate $\gamma_{{\rm non\text{-}osc},\,j:(j-k)}$, together with the equilibrium values of $e_{2, \rm eq}$, $\phi_{2, \rm eq}$, and $\kappa_{2,\rm eq}$, for different MMRs are summarized in Table~\ref{tab:mmr_eq}. We compare the analytic growth rates with the numerical linearization results in Figure~\ref{fig:growthrate}, finding excellent agreement over a large range in $\beta$ and for several low-order resonances. We also compare these semi-analytic growth rates with numerical EEOM orbit integrations in Figure~\ref{fig:odeplot}. The bottom panel shows the growth of the libration amplitude, with the semi-analytic growth rate in Equation \ref{eq:gamma_formula} giving good agreement.

The perturbative correction $s_1$ has a simple physical interpretation. The PR drag terms provide an additional ``force" $A_1 r$ acting on the particle, and the projection of that force back on the particle's motion, $l^T A_1 r$, is the ``force dot velocity" work done. For the $\omega=\pm \omega_0$ eigenfunctions, since they have no $r$ component along $\delta \kappa_2$, only the $\dot{\Gamma}_{,\Gamma}$ and $\dot{\kappa}_{,\Gamma}$ terms in $A_1$ can contribute. These terms correspond to changes in $\dot{\Gamma}_2$ and $\dot{\kappa_2}$ in response to variations in $e_2$ along the trajectory. For these extra ``forces" to drive growth, they must have the right phase relative the unperturbed motion. By contrast, for the $\omega=0$ mode, $l$ only has a component along $\delta \kappa_2$. Therefore only the $\dot{\kappa}_{,\Gamma}$ and $\dot{\kappa}_{,\kappa}$ terms can contribute to the eigenvalue correction.

\begin{figure*}[ht!]
    \centering
    \includegraphics[width=\textwidth]{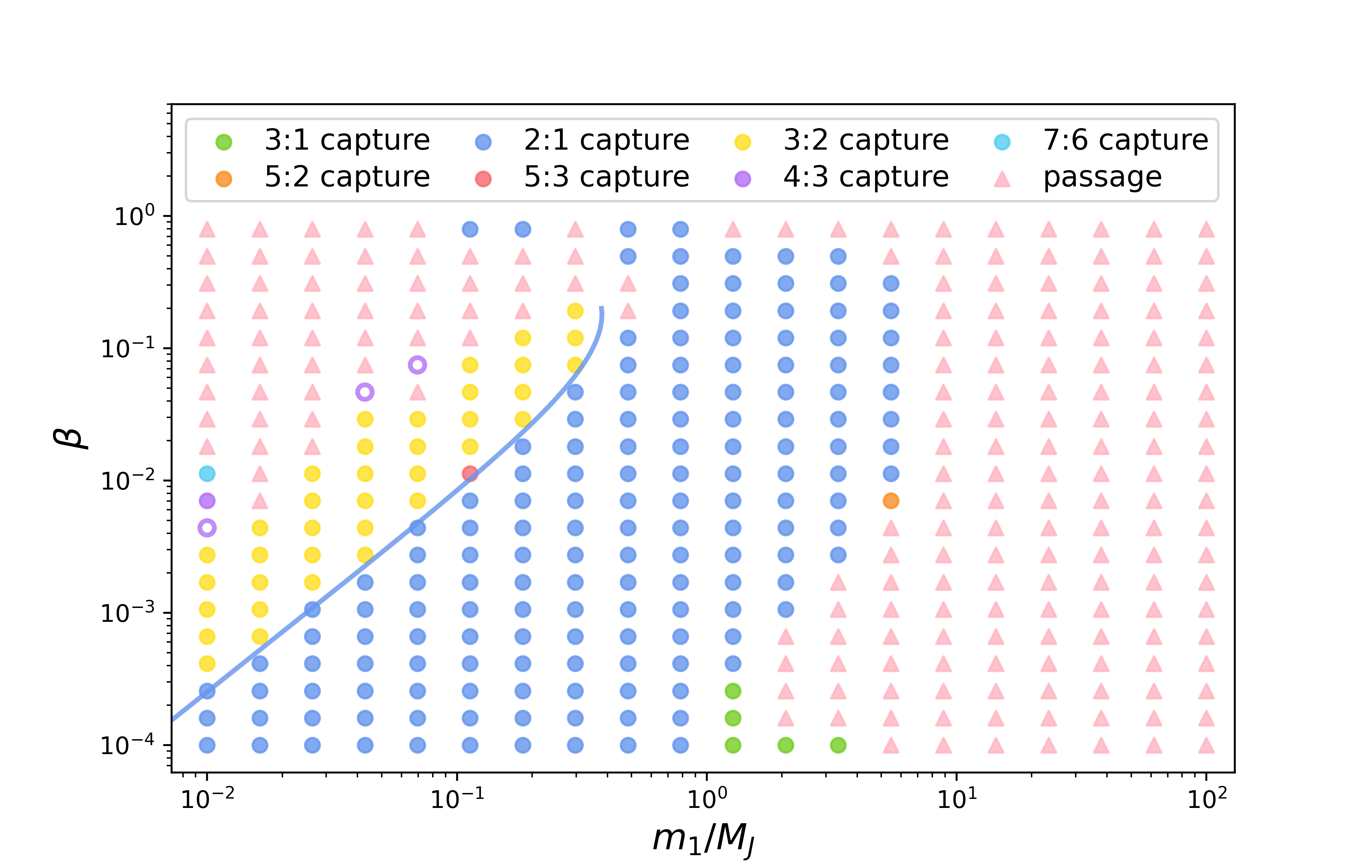}
    \caption{Capture and passage outcomes in the $(m_1,\beta)$ parameter space for $m_0=1\,M_\odot$ and $a_1\simeq0.05\,{\rm AU}$. Circles denote capture and triangles denote passage. Filled symbols indicate that all initial phases give the same outcome; open symbols indicate mixed outcomes, with the color showing the most frequent outcome. The blue curve shows the adiabatic capture boundary for the $2\!:\!1$ MMR.
    \label{fig:captureplot}}
\end{figure*}

\section{Capture into Resonance}\label{sec5}

In our orbital integrations, the dust particle was launched at wide separation, $a_2 = 5a_1$, and several different low-order resonances may be encountered as the orbit decays due to PR drag. In practice, however, the dust particle is typically captured into at most one resonance.
The goal of this section is to determine which resonances are important, 
as a function of $\beta$ and planet mass $m_1$, and to compare to analytic expectations.

Here we define ``capture into resonance'' as capture into a dissipative equilibrium. For each simulation, we inspect plots of $a_2$, $e_2$, and $\phi_2$ versus time, and declare a particle as captured when each quantity showed small oscillation around $a_{\rm 2, eq}$, $e_{\rm 2, eq}$, and $\phi_{\rm 2, eq}$ (see Table \ref{tab:mmr_eq}). Particles that temporarily librate near the initial conservative fixed points but leave the resonance before reaching the dissipative equilibrium are instead classified as passage. 

To map the capture outcomes, we use \textsc{Rebound} to perform EEOM orbit integrations for a Sun-like host star ($m_0 = 1 M_\odot$) and a planet at semi-major axis $a_1 \simeq 0.05\,\mathrm{AU}$. Simulations are carried out for  20 logarithmically spaced values of $\beta$ from $10^{-4}$ to $10^{-0.1}$, and 20 logarithmically spaced values of $m_1$ from $10^{-2}$ to $10^{2}\,M_J$. For each point in this parameter grid, we perform six integrations with identical initial conditions but different initial values of $\lambda_2$ uniformly spaced in $[0,2\pi)$. In most cases, the outcomes are consistent across these realizations and are assigned to each point, indicated by different colors in the figure. When the outcomes differ for different initial angles, the most probable outcome is shown using hollow circles.

Figure~\ref{fig:captureplot} shows the resulting capture map in the $\beta$--$m_1$ plane (e.g., \citealt{2023ApJ...946L..11B}), with different symbol shapes and colors for various outcomes. Capture is favored at smaller $\beta$, and hence slower migration through resonance. The 2:1 and 3:2 resonances dominate capture over the parameter space. In the upper left corner, at small $m_1$ and large $\beta$, the migration rate is so high that particles cross resonances without reaching $e_{\rm 2, eq}$. Along the middle and lower left side, particles can pass through the outer resonances (e.g., 3:1, 2:1) before being captured into an inner resonance (e.g., 3:2, 4:3). In the middle of the plot, a broad swath of $\beta-m_1$ space is dominated by capture into  2:1 resonance. 

One might have expected that on the right side of the plot, at large $m_1$, capture would be easier as the gravity of the planet is stronger. These cases do show initial libration around a conservative equilibrium point, but they then suddenly exit resonance before reaching the dissipative fixed point. We suspect that the Q-averaged $\langle R_2 \rangle$ may be insufficient to understand capture into MMR for such high planet masses, and other terms in $R_2$ may become important at large $m_1$.

We now interpret these outcomes using an adiabatic capture criterion, focusing on capture into the 2:1 MMR. Capture is expected if the eccentricity of the dust particle far from resonance is sufficiently small and the migration rate is sufficiently slow compared to the libration frequency around the equilibrium point \citep{2014AJ....147...32G}. As we initialize particles with $e_2=0$, and PR drag damps the eccentricity during the inward migration, the first condition is satisfied. The relevant question is therefore whether the resonant fixed point evolves slowly enough for the particle to follow it adiabatically.

The \citet{2014AJ....147...32G} argument uses the ${\cal O}(e_2)$ LPE (see Appendix \ref{app:firstorder}). The minimum libration frequency of the incoming equilibrium point at $\phi_{\rm 2, eq}=\pi$ is $\omega_{\rm 0, min} \sim n_2 (\mu_1 C_1)^{2/3}$ , at which point $e_2 \simeq (\mu_1 C_1)^{1/3}$. At small $e_2$, $\kappa_2(t) \simeq -2 t/t_{\rm pr} \sim - e_2^2$ (see Equation \ref{eq:kappa2dot}) and so the time to reach an eccentricity $e_2 \sim (\mu_1 C_1)^{1/3}$ is
$t \sim t_{\rm pr} (\mu_1 C_1)^{2/3}$. 
For this time to be longer than the libration time $\omega_{\rm 0, min}^{-1}$
then requires $n_2 t_{\rm pr} (\mu_1 C_1)^{4/3} \ga 2.5$.

This criterion is shown as the blue curve in Figure~\ref{fig:captureplot}, and performs well for most of the 2:1 MMR cases. This is perhaps surprising, as inclusion of ${\cal O}(e_2^2)$ terms in $R_2$ causes the libration frequency to decrease monotonically to zero (for the $m_1/m_0$ values of interest here) toward the bifurcation at $e_2 \simeq 0.03$, beyond which the $\phi_2=\pi$ equilibrium point becomes unstable. So there is no minimum libration frequency in the true dynamics. Nevertheless, this condition seems to work well to understand not only capture into conservative but also dissipative equilibrium points.

\section{Stochastic Evolution After Escape from Resonance }\label{sec6}

\begin{figure}[ht!]
    \centering
    \includegraphics[width=\columnwidth]{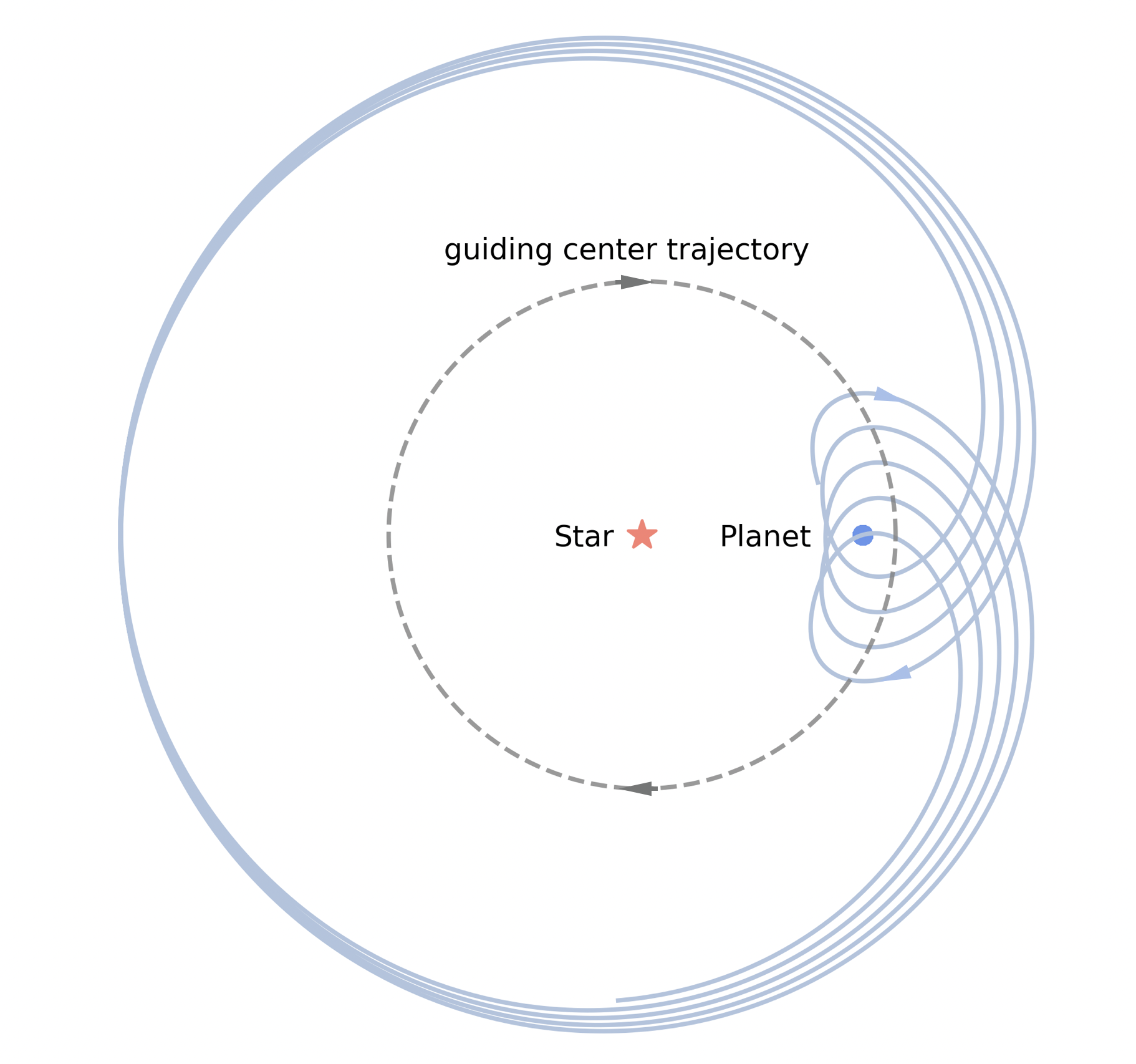}
    \caption{A test particle moves as an epicycle of radius $\rho=a_2e_2$ with guiding center circulating around the planet with radius $a_2$ in the rotating frame. The example shown uses Keplerian orbits with $a_2=1.58\,a_1$,  $e_2=0.48$, $i_2=0$, and $\varpi_2=0.29~{\rm rad}$.}
    \label{fig:epicycle}
\end{figure}

In previous sections, we demonstrated that resonances are short lived:
overstability of the dissipative equilibrium points eventually drives the particle across the separatrix, leading to escape from resonance. Due to the high $e_{\rm  2, eq}$ in resonance, the particle and planet exit the resonance on crossing orbits, and close encounters become possible near conjunctions. Figure \ref{fig:epicycle} shows a Keplerian orbit of the particle in the frame rotating with the planet. The ``loop" feature rotates around when out of resonance, and close approaches occur when the loop sweeps by the planet. 

\subsection{Modeling the Effect of Close Encounters} 

For close encounters, the particle's orbit will receive an impulsive perturbation which will cause the orbit to change in a stochastic manner in time. This diffusion causes orbital energy $\varepsilon =  - Gm_0(1-\beta)/(2a_1) \times x$, to vary, where $x=a_1/a_2$ is a convenient dimensionless variable representing orbital energy. Eccentricity $e_2$ also varies, as the Jacobi constant $C_J$ is unchanged by close encounters, and only changes due to PR drag on longer timescales. These random changes are called ``orbital diffusion" in the context of comets scattered by the giant planets \citep{1980A&A....85...77Y}.

Initially $x$ is given by the resonant value $x_{\rm res} \simeq a_1/a_{\rm 2, res}$ as the resonance is exited. Kicks $\Delta x$ occur each orbit, causing $x \rightarrow x + \Delta x$, 
with a distribution $P(\Delta x)$ calculated below. 
The resultant random walk in $x$ has absorbing boundaries at $x_{\rm ej}=0$ and $x_{\rm coll, s}=x_{\rm res}$ for ejection and hitting the sublimation radius of the star, respectively. PR drag drives the orbit to decay to larger $x$, while orbital diffusion can either increase or decrease $x$. Collisions with the planet have a per-orbit probability $p_{\rm coll, p}$. Depending on the orbital diffusion and PR decay timescales, it becomes likely to hit the planet when $\sim p_{\rm coll, p}^{-1}$ orbits have occurred. 

\begin{figure*}[ht!]
    \centering
    \includegraphics[width=\textwidth]{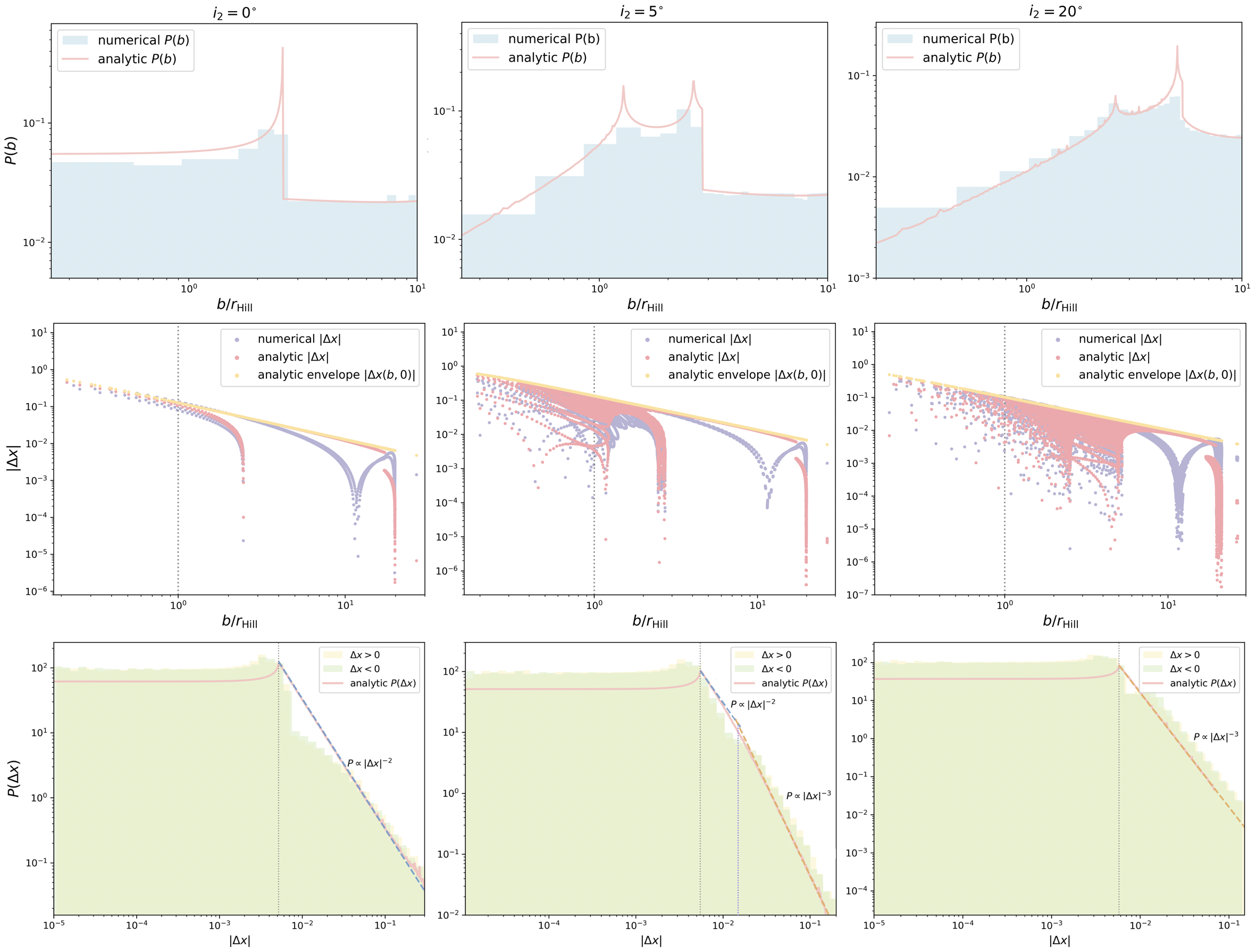}
    \caption{Distribution of impact parameters $P(b)$, energy changes $\Delta x$ versus impact parameter, and distribution of energy changes for Keplerian orbits which are crossing. 
    Top: Numerical (blue histograms) and analytic (red curves) impact parameter distributions $P(b)$ for inclinations $i_2=0^\circ,\,5^\circ$, and $20^\circ$. Middle: Numerical $|\Delta x|$ versus $b/r_{\rm Hill}$ (purple), together with the analytic $|\Delta x|$ (red) and envelope $|\Delta x(b, 0)|$ (yellow), which provides an excellent upper bound on the numerical results. Bottom: Numerical energy change distributions $P(\Delta x)$, with positive (yellow) and negative (green) $\Delta x$ shown separately, compared with the analytic $P(\Delta x)$ (red) of a transition from a $|\Delta x|^{-2}$ (blue dashed lines) to a $|\Delta x|^{-3}$ (orange dashed lines) power-law tail. The gray dotted line marks the transition kick size $\Delta x_{\rm trans}$ between small and large kicks (Equation \ref{eq:deltax_trans}), and the purple dotted line marks the critical kick size $\Delta x_{\rm cr}$ (Equation \ref{eq:deltax_crit}).}
    \label{fig:P_b_P_dx}
\end{figure*}

To understand $P(\Delta x)$, we perform orbit integrations of a test particle around the star--planet binary, exploring the range of possible orbital orientations and phases.
For each set of fixed initial parameters $(a_1,a_2,e_2,i_2)$, single-orbit integrations are carried out for different values of $M_1$, $\omega_2$, and $\Omega_2$. The particle is started at apocenter, and the integration time is one (initial) orbital period. Treating $(M_1,\omega_2,\Omega_2)$ as uniformly distributed over $[0,2\pi)$, the integrations give rise to a distribution $P(\Delta x;a_1,a_2,e_2,i_2)$ for $\Delta x=a_1/a_{\rm 2, final}-a_1/a_{\rm 2, initial}$, where initial and final refer to the start and end of the integration, respectively. The possibility of collisions with the planet and star sublimation zone is included. Integrations where a collision occurs do not produce an $a_{\rm 2, final}$, and hence do not contribute to $P(\Delta x)$. Ejections can also occur, and the corresponding $\Delta x$ in these integrations is included in $P(\Delta x)$. The position and velocity of the star and planet are tabulated over the orbit, and the closest approach is used to define the impact parameter $\mathbf{b}$ and relative velocity $\mathbf{u}$ \footnote{As we use the closest approach of the exact orbit, these are not ``at infinity" quantities. A better method may have been to use non-interacting, Keplerian orbits. However, most collisions have impact parameter larger than the 90 degree scattering length for the parameters assumed here, and so the corrections are not expected to be large.}. As more than one local closest approach may occur per orbit, we report only the smallest $b$ solution.

The \texttt{REBOUND} code \citep{2012A&A...537A.128R} is used to carry out the single-orbit integrations. Radiation forces are ignored in these integrations ($\beta=0$). For each of $M_1$, $\omega_2$, and $\Omega_2$, we use $1024$ grid points, yielding a total of $2^{30}$ runs. As close encounters only occur for specific choices of the parameters, filling out the distributions to large $\Delta x$ required many points when uniform gridding is used. We adopt $m_0=1 M_\odot$, $m_1=1 M_J$, and $a_1\simeq 0.05\,{\rm AU}$. For the physical radii, we use $R_{\rm sub}\simeq 5.85\,R_\odot$ and $R_1=1\,R_J$. The test particle is initialized at the resonant values for the $\!2:\!1$ resonance, with $a_2=2^{2/3}a_1$ and $e_2\simeq 0.4812$.

First we focus on the distribution of impact parameters, $P(b)$, shown in the top three panels of Figure \ref{fig:P_b_P_dx}. A naive expectation for $P(b)$ would be $P(b)db \propto 2\pi b db$, i.e., uniform in area around the target. The left panel shows $P(b)$ for $i_2=0^\circ$. At small $b$ for close encounters, $P(b)$ is nearly {\it flat} instead of rising linearly in $b$, a result of the coplanar encounters. At larger $b$, comparable to typical orbital separations, the distribution exhibits a ``spike,'' followed by a nearly flat distribution at the largest $b$. By contrast, the $i_2=5^\circ$ and $20^\circ$ cases find the expected $P(b) \propto b$ at small $b$, with a spike and transition to a nearly constant distribution, and then another spike and flat distribution at large $b$.

To understand the impact parameter distribution, a toy model is developed in Appendix \ref{app:impact_parameter}. The particle's orbit is represented by a circular epicycle of radius $\rho=a_2e_2$ moving around on a guiding center circle of radius $a_2$, as in Figure \ref{fig:epicycle}. At each position of the guiding center, $b$ is defined to be the distance from the planet to the closest point on the epicycle. 
The distribution $P(b)$ for close approaches has the analytic form
\begin{equation}\label{eq:P_small_b}
P(b)
\simeq  C_ 0\frac{b}{\sqrt{b^2 + (a_1\sin{i_2})^2}}
\end{equation}
where the normalization constant is
\begin{equation}
C_0 =  \frac{4a_2e_2}{\pi\sqrt{\big((a_2+a_1)^2-(a_2e_2)^2\big)\big((a_2e_2)^2-(a_2-a_1)^2\big)}}.
\end{equation}
The denominator in Equation \eqref{eq:P_small_b} shows that when $a_1\sin i_2 \gg b$, the distribution is uniform in area, 
while for $a_1 \sin i_2 \ll b$ it approaches a  plateau at small $b$. 

Figure~\ref{fig:P_b_P_dx} (upper panels) compares the model predictions (red curves; Equation \ref{eq:P_of_b_inclined}) with numerical orbital integrations (blue histograms), showing qualitative and some quantitative agreement for both coplanar and  inclined cases. At large $b$, the model reproduces the spike that occurs when the guiding center is opposite the planet, as well as the flat distribution at typical orbital separations. For small $b$, the model exhibits the same behavior found in the numerical integrations: a flat distribution in the coplanar case and a distribution that is uniform in area in the inclined case.

Next, we turn to the per-orbit energy changes $\Delta x$ shown in the middle three panels of Figure \ref{fig:P_b_P_dx}. The numerical $\Delta x$ results (blue points) are from numerical integrations, using the initial and final $a_2$. Most of the impact parameter range has $b \gg b_0$ and so $\Delta x \propto b^{-1}$, which sets the observed slope. For the two inclined cases, a spread is seen below the upper envelope, and less spread is seen for the coplanar case. There are also dips observed at certain $b$ where $\Delta x \rightarrow 0$. Aside from the dip at large $b$, the $b^{-1}$ scaling works surprisingly well even outside the Hill sphere, where non-Keplerian effects such as stellar tidal forces are expected to significantly change the answer \citep{1986CeMec..38...67H}. For orbits with distant encounters, short period terms in the disturbing function might be expected to produce changes $\Delta x \sim m_1/m_0 \sim 10^{-3}$.

To understand the results for $\Delta x$ from numerical integrations, 
an analytic expression for $\Delta x$ is developed in Appendix~\ref{app:pdx}, 
treating both the test particle and the planet as moving on Keplerian orbits.
The ``kick" $\Delta x$ from a close approach depends on the impact parameter $b$ and the orientation angle $\chi$ where $\cos \chi = \mathbf{v}_1 \cdot \mathbf{b}/(b v_1)$ and $\mathbf{v}_1$ is the velocity of the planet around the star at the time of closest approach. The result given in Equation \eqref{eq:dx_of_b} is
\begin{equation} \label{eq:dx_of_b_chi}
\Delta x(b,\chi)
= \left( \frac{2 \sqrt{3 - 2\beta - C_J}}{1-\beta} \right)\, \left( \frac{2(b/b_0)}{1+(b/b_0)^2} \right) \cos\chi,
\end{equation}
where the 90 degree scattering length is $b_0=Gm_1/u^2 \simeq a_1 (m_1/m_0)/(3 - 2\beta - C_J)$. For $b \gg b_0$, this expression gives $\Delta x \propto b^{-1}$. If $\chi$ is uniformly distributed, $\cos \chi$ gives alternating signs as well as the spread below the maximum value $\Delta x_{\rm max}=\Delta x(b,0)$. The dips where $\Delta x \rightarrow 0$ occur at certain $b$ where $\mathbf{b}$ and $\mathbf{v}_1$ are nearly perpendicular. In addition to computing $\mathbf{b}$, the numerical integrations also give $\chi$. Histograms of $\chi$ (not shown) indicate an approximately uniform distribution for the inclined cases. For the $i_2=0^\circ$ case, less scatter is seen in $\Delta x$ since $\chi$ has two broad peaks instead of a uniform distribution. However, in the absence of a simple scheme to compute these $\chi$ distributions, we will later assume uniformly distributed $\chi$ in the Monte Carlo simulations below.

The three middle panels of Figure \ref{fig:P_b_P_dx} compare the numerical results (blue dots) against the analytic formula (Equation \ref{eq:dx_of_b_chi}), using values of $b$ and $\chi$ determined from the numerical integration. Note that we do not use the measured $\mathbf{u}$ at closest approach, but rather the approximation $u_\infty=\sqrt{Gm_0/a_1}(3-2\beta-C_J)^{1/2}$, valid for close encounters. Equation \eqref{eq:dx_of_b_chi} shows good agreement for $b \la (2$--$5) \times r_{\rm Hill}$, showing the expected slope and scatter, as well as the dips where $\cos \chi \simeq 0$. This agreement is perhaps better than would have been naively expected so far outside the Hill radius. At larger impact parameters, $b \ga 10\, r_{\rm Hill}$, the numerical integrations show a dip not present in the analytic formula.

Next we consider the distribution of energy changes $\Delta x$. The yellow and green shaded regions in the bottom panels of Figure~\ref{fig:P_b_P_dx} show $P(\Delta x)$ as determined from numerical simulations. For distant encounters (small $\Delta x$) the distribution is flat, changing into a power-law $P(\Delta x) \propto \Delta x^{-\alpha}$ for $\Delta x \ga m_1/m_0 \sim 10^{-3}$. A  ``cusp plus notch" feature is seen near this transition. In the coplanar case, or the inclined case at smaller $\Delta x$, the power-law index is $\alpha \simeq 2$, while for the inclined case at larger $\Delta x$ it steepens to $\alpha \simeq 3$. This $\alpha=3$ power-law has been reported in previous studies \citep[e.g.,][]{1968AJ.....73.1039E,1987AJ.....94.1330D} .

Given the small-$b$ impact parameter distribution (Equation \ref{eq:P_small_b}), the uniform orientation angle distribution $P(\chi)d\chi = d\chi/\pi$, and the kick formula $\Delta x(b,\chi)$ in Equation \eqref{eq:dx_of_b_chi}, a change of variables gives the distribution $P(\Delta x)$. The derivation is given in Appendix~\ref{app:pdx}, with the result
\begin{equation}\label{eq:p_dx_large_unif}
P(\Delta x)
\simeq
C_0\,
\frac{(2b_0\Delta x_0)^2}
{|\Delta x|^2\sqrt{(2b_0\Delta x_0)^2 + (a_1\sin i_2)^2 |\Delta x|^2}}
\end{equation}
in the power-law region. Here $\Delta x_0=2\sqrt{3-2\beta-C_J}/(1-\beta)$ is the maximum kick size, attained for $b=b_0$ and $\cos{\chi}=1$. The denominator gives a critical kick size
\begin{equation}\label{eq:deltax_crit}
\Delta x_{\rm cr} \simeq \frac{2b_0\Delta x_0}{a_1 \sin i_2},
\end{equation}
such that $\alpha=2$ for $|\Delta x| \la \Delta x_{\rm cr}$ and $\alpha=3$ for $|\Delta x| \ga \Delta x_{\rm cr}$. The origin of these slopes can be understood simply. Since $\Delta x\propto b^{-1}$, we have $db/d\Delta x\propto (\Delta x)^{-2}$, and therefore
\begin{equation}
P(\Delta x)\sim \frac{P(b)}{(\Delta x)^2}.
\end{equation}
For $P(b)\sim {\rm constant}$, $P(\Delta x)\propto (\Delta x)^{-2}$, while for $P(b)\propto b \propto (\Delta x)^{-1}$, $P(\Delta x)\propto (\Delta x)^{-3}$. Thus, as $i_2$ decreases, a larger fraction of the power-law region follows $P(\Delta x)\propto (\Delta x)^{-2}$, whereas increasing $i_2$ shifts more of the power-law region toward $P(\Delta x)\propto (\Delta x)^{-3}$.

The bottom three panels of Figure \ref{fig:P_b_P_dx} compare the numerical integrations against the numerical integral for $P(\Delta x)$ from Equation \eqref{eq:p_dx_unif}. This integral covers the full range of $\Delta x$ and captures the cusp feature near the transition from the flat part of the distribution to the power-law tail. The analytic results show good qualitative, and some quantitative agreement with the simulations. The $i_2=0^\circ$ case agrees with the expected $\alpha=2$ power-law, while the $i_2=20^\circ$ case agrees with $\alpha=3$ over most of the power-law region. The $i_2=5^\circ$ case transitions between $\alpha=2$ and $3$ regions, although this behavior is partly obscured by the presence of the notch. The analytic result overestimates in the notch region, and therefore underestimates in both the flat and power-law regions.

Some previous treatments have assumed $P(\Delta x)$ is Gaussian for simplicity, as this allows the time-dependent distribution function $P(x,t)$ to be computed from the Fokker-Planck equation \citep{1980A&A....85...77Y}. However, both the numerical integrations and the analytic solution in Equation \eqref{eq:p_dx_large_unif} find a flat distribution $P(\Delta x)$ at small $\Delta x$, and a power-law $(\Delta x)^{-\alpha}$ at large $\Delta x$, which is 
only roughly approximated by the first two moments of $\Delta x$. 

\subsection{Monte Carlo Evolution of Post-Resonance Orbits}
\label{sec6.4}

Given the agreement between the orbit integrations and analytic approximations for $P(b)$, $\Delta x(b,\chi)$, and $P(\Delta x)$ , the next section describes their implementation in a Monte Carlo algorithm for post-resonance orbital evolution. 
This approach is much faster than direct integrations including PR drag with \texttt{REBOUND} and \texttt{REBOUNDx}, and allows us to assess the importance of orbital diffusion, PR drag, ejections, and collisions with the star and planet.

The procedure is implemented as follows. Each particle is first
initialized at the MMR equilibrium state $(e_{2,\rm eq},a_{2,\rm eq})$, after which a loop is entered with the following steps.
(i) The orbit-averaged PR drag terms in Eqs.~\eqref{eq:dadt} and \eqref{eq:dedt} are used to update $a_2$ and $e_2$ over one orbital period. (ii) We check whether the pericenter distance satisfies
$a_2(1+e_2)<a_1$, implying that it is no longer orbit crossing and will undergo PR drag into the star. When this condition is triggered, the subsequent orbital decay is skipped and the final fate is recorded as
\emph{collision with the star (sublimation)}, terminating the evolution. (iii) Otherwise, we compute the Jacobi constant $C_J(a_2, e_2)$, which is conserved during the encounter. For inclined orbits, this Jacobi constant is given by
\begin{equation}\label{eq:CJ_inc}
\begin{aligned}
\frac{Gm_0}{2a_1}C_J = &-\frac{Gm_0(1-\beta)}{2a_2}\\[4pt]
&- n_1\sqrt{Gm_0(1-\beta)a_2(1-e_2)^2}\cos{i_2},
\end{aligned}
\end{equation}
where $i_2$ is held fixed for simplicity.
The outcome of the encounter is determined by the analytic probabilities for planetary collision
\begin{equation}
\begin{aligned}\label{eq:p_col}
p_{\rm p, coll}
&=\int_{0}^{R_{\rm gf}} 
C_0\,\frac{b}{\sqrt{b^2+(a_1\sin i_2)^2}}\,db \\[4pt]
&= C_0\left(\sqrt{R_{\rm gf}^2+(a_1\sin i)^2}-a_1\sin i_2\right),
\end{aligned}
\end{equation}
and for entry into the Hill sphere without planetary collision
\begin{equation}\label{eq:p_Hill}
\begin{aligned}
p_{\rm Hill}
&=\int_{R_{\rm gf}}^{r_{\rm Hill}} C_0\,\frac{b}{\sqrt{b^2+(a_1\sin i_2)^2}}\,db \\[4pt]
&= C_0\Bigl(\sqrt{r_{\rm Hill}^2+(a_1\sin i_2)^2}
-\sqrt{R_{\rm gf}^2+(a_1\sin i_2)^2}\Bigr),
\end{aligned}
\end{equation}
where $R_{\rm gf}=R_1\sqrt{1+2Gm_1/(R_1 u_\infty^2)}$ is the gravitationally focused radius. If a random number $p_{\rm rand}<p_{\rm coll, p}$, the fate is recorded as \emph{collision with the planet}. If instead $p_{\rm coll,p}<p_{\rm rand}<p_{\rm coll,p}+p_{\rm Hill}$, the particle will receive a kick $\Delta x$ from an encounter inside the Hill sphere. In this case we sample the impact parameter $b$ from Equation~\eqref{eq:P_small_b}, draw $\chi$ uniformly in $[0,\pi)$, and compute $\Delta x(b,\chi)$ from 
Equation~\eqref{eq:dx_of_b}. Finally, if $p_{\rm rand}>p_{\rm coll,p}+p_{\rm Hill}$, the encounter occurs outside the Hill sphere, where the $\Delta x$ distribution is observed to be roughly flat in Figure \ref{fig:P_b_P_dx}. We model the corresponding energy kick 
$\Delta x=\xi (m_1/m_0)$, where $\xi$ is uniform in $[-1,1]$.
Given $\Delta x$, the post-encounter semi-major axis is updated according to $a_{2,\rm after}
=a_1/\left(a_1/a_{2,\rm before}+\Delta x\right)$.
(iv) If $a_{2,\rm after}<0$, the particle's fate is recorded as \emph{ejected}. (v) Otherwise, the eccentricity $e_{2,\rm after}$ is obtained by the conserved Jacobi constant $C_J(a_2,e_2)$, using the updated $a_{2,\rm after}$. We then check whether the new pericenter satisfies $a_{2,\rm after}(1-e_{2,\rm after})<R_{\rm sub}$, where $R_{\rm sub}$ is the sublimation radius. If so, the final fate is
\emph{collision with the star (sublimation)}. (vi) If not, the integration time is advanced by one orbital period, $t_{\rm after}=t_{\rm before}+P_{\rm dust}$, and the procedure is repeated until the particle reaches one of the three fates.

\begin{figure*}[ht!]
    \centering
    \includegraphics[width=\textwidth]{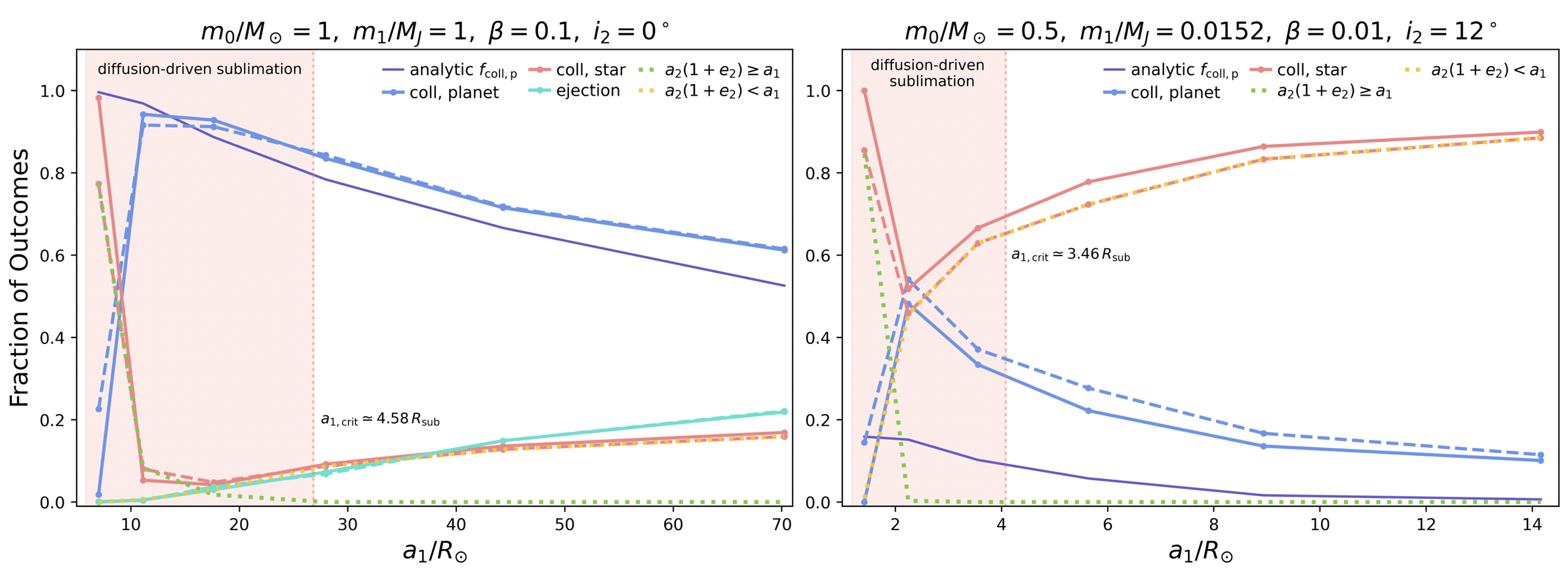}
    \caption{Fraction of outcomes as a function of $a_1/R_\odot$ comparing Monte Carlo results (solid) with EEOM \textsc{Rebound} integrations (dashed). The red solid line, for hitting the star, is broken down into two contributions, for systems with $a_2(1+e_2) \geq a_1$ (green dotted; ``diffusion-aided" collision) and those with $a_2(1+e_2) < a_1$ (yellow dotted; ``PR decay" collision") at the time of collision. Diffusion-driven collision with the star is possible in the shaded region with 
    $a_1 < a_{1,\rm crit}$, indicated by a vertical red dotted line.
    The analytic estimate for collision with the planet
    (Equation \ref{eq:f_col}) is also shown (purple). Left: Coplanar case ($i_2=0^{\circ}$) with $m_0/M_{\odot}=1$, $m_1/M_J=1$, and $\beta=0.1$, corresponding to a dust radius of $r_d \simeq 2.3\,\mathrm{\mu m}$. Right: Inclined case ($i=12^{\circ}$ with $m_0/M_{\odot}=0.5$, $m_1/M_J=0.015$, and $\beta=0.01$, corresponding to a dust radius of $r_d \simeq 1.89\,\mathrm{\mu m}$.
    \label{fig:fraction}}
\end{figure*}

We apply this Monte Carlo scheme to ensembles of $N=10^4$ evolutions for two representative planetary systems. In Figure \ref{fig:fraction}, the results are compared to direct EEOM orbital integrations using \texttt{REBOUND} and \texttt{REBOUNDx}. Each panel shows the fractions of dust particles that collide with the planet, collide with the star, or become ejected as a function of the planet’s orbital radius $a_1$. Solid curves denote the Monte Carlo results, while dashed curves show the corresponding fractions from EEOM numerical integrations. For our standard case of dust entering the $2\!:\!1$ resonance around a Sun-like star ($m_0=1\, M_{\odot}$) and a Jupiter-mass planet ($m_1=1\, M_J$), with $i_2=0$ and $\beta=0.1$, the fraction of collision is very high at small separation, and gradually decreases as $a_1$ increases. In contrast, for dust entering 3:2 resonance around an M-dwarf star ($m_0=0.5\,M_\odot$) and a sub-Neptune planet ($m_1=0.015\,M_J$), with $i_2=12^\circ$ and $\beta=0.01$, the fraction of collision is again high at small $a_1$ but declines more rapidly. In this regime, hitting the star becomes the dominant outcome at larger orbital separation. The agreement is good in both cases, demonstrating that the Monte Carlo model 
approximates the post-resonant evolution well.

There are two distinct channels leading to sublimation near the star. In the first, the particle has exited the orbit-crossing region with the planet, $a_2(1+e_2) < a_1$, and subsequently migrates inward under PR drag until sublimation. In the second, the particle remains orbit-crossing, $a_2(1+e_2) \geq a_1$, but reaches sufficiently high eccentricity that its pericenter falls below the sublimation radius, $a_2(1-e_2) < R_{\rm sub}$. These two contributions are shown separately in Figure~\ref{fig:fraction} (yellow and green dotted lines). The second channel can be understood in the $(a_2, e_2)$ plane. For a fixed $C_J(a_2, e_2)$, orbits lie along a constant-$C_J$ curve. The orbit-crossing boundary is $a_2(1+e_2)=a_1$, while sublimation requires $a_2(1-e_2)=R_{\rm sub}$. Their intersection sets the minimum pericenter,
and thus a critical ratio $a_{1,\rm crit}/R_{\rm sub}$. If $a_1<a_{1,\rm crit}$, there exists a segment of the constant-$C_J$ curve satisfying both conditions, allowing diffusion-driven sublimation. For the $2\!:\!1$ MMR, $a_{1,\rm crit} \simeq 4.58\,R_{\rm sub}$ (left panel), and for the $3\!:\!2$ MMR, $a_{1,\rm crit} \simeq 3.46\,R_{\rm sub}$ (right panel). For $a_1 < a_{1,\rm crit}$, sublimation is dominated by this channel. Compared with \citet{2018MNRAS.480.5560B} (hereafter Bonsor18), the fraction of collision with the planet does not increase inward monotonically, but instead declines at small separations, where diffusion-driven sublimation becomes efficient, reflecting the inclusion of this additional pathway.

To better understand the fraction hitting the planet, we analytically estimate this quantity using the single-orbit collision probability with the planet, $p_{\rm coll, p}$, and the number of planet-cross orbits under PR-driven decay, $N_{\rm pr}$ as
\begin{equation}\label{eq:f_col}
f_{\rm coll, p}=1-(1-p_{\rm coll,p})^{N_{\rm pr}}
\simeq 1 - e^{-N_{\rm pr}p_{\rm coll,p}},
\end{equation}
where the second expression is valid when $N_{\rm pr}p_{\rm coll,p}^2 \ll 1$.
To give a simple estimate of the collision probability, we ignore gravitational focusing, treat $a_1 \sin i_2 \gg R_1$, and use $2\!:\!1$ resonance values for $a_{\rm 2, res}$ and $e_{\rm 2, res}$ to find
\begin{equation}
p_{\rm coll, p} \simeq
2.3 \times 10^{-4} \left( \frac{R_1}{R_J} \frac{10\, R_\odot}{a_1} \right)^2 \left(\frac{10^\circ}{i_2} \right).
\end{equation}
The number of planet-crossing orbits  is obtained by integrating the PR drag terms in Eqs. \eqref{eq:dadt} and \eqref{eq:dedt}, from the resonant eccentricity $e_{\rm 2, res}$ until the particle detaches from the planet orbit-crossing region at $a_{2,\rm end}(1+e_{2,\rm end})<a_1$. Using the PR drag constant of motion $C_{\rm pr}=a_2(1-e_2^2)e_2^{-4/5}$ to express $a_2=a_2(e_2)$, the integral can be evaluated analytically:
\begin{equation} \label{eq:Npr}
\begin{aligned}
N_{\rm pr}
&
= \int_{e_{2,\rm res}}^{e_{2,\rm end}}
\frac{1}{P_2\!\bigl(a_2(e_2)\bigr)}
\frac{dt}{de_2}\,de_2 \\[4pt]
&\simeq 350\,
\frac{\sqrt{1-\beta}}{\beta}
\left(\frac{j}{j-k}\right)^{1/3}
\left(\frac{M_\odot}{m_0}\right)^{1/2}
\left(\frac{a_1}{10\, R_\odot}\right)^{1/2} \\
&\quad \times
\sqrt{1-e_{2,\rm res}^2} \left[1-\left(\frac{e_{2,\rm end}}{e_{2,\rm res}}\right)^{2/5}\right].
\end{aligned}
\end{equation}
Evaluating the product for the $\!2:\!1$ resonance gives the approximate collision fraction
\begin{eqnarray}\label{eq:f_col_p}
    f_{\rm coll, p} & \simeq & N_{pr}p_{\rm coll, p} 
    \simeq 0.2\,
\left( \frac{0.1\sqrt{1-\beta}}{\beta} \right)
\left(\frac{M_\odot}{m_0}\right)^{1/2} \left( \frac{R_1}{R_J} \right)^2
\nonumber \\ & \times & \left(\frac{a_1}{10\, R_\odot}\right)^{-3/2} 
\left(\frac{10^\circ}{i_2} \right),
\end{eqnarray}
which shows an order unity size, and increases for larger planets, smaller orbital separations, and lower inclinations.

Figure \ref{fig:fraction} shows the analytic estimate for collision with the planet (Equation \ref{eq:f_col}) evaluated using the exact expressions for $N_{\rm pr}$ (Equation \ref{eq:Npr}) and $p_{\rm coll,p}$ (Equation \ref{eq:p_col}). In the left panel ($i_2=0^{\circ}$, $2\!:\!1$ MMR), the $f_{\rm coll,p}$ estimate captures the correct qualitative behavior, but  underestimates the numerical results, perhaps due to the neglect of orbital diffusion. In the right panel ($i_2=12^\circ$, 3:2 MMR), the analytic formula shows a larger underprediction of planetary collisions.

For further comparison, Bonsor18 carry out similar orbit integrations including PR drag as in our work. One difference is that their semi-major axes are sufficiently far from the star that the ``diffusion-aided" hitting of the sublimation zone around the star does not occur; instead particles are only removed through the ``detach from the planet and decay down to the star" channel. We have compared a handful of cases to the results reported in their Table A5 and find agreement between numerical integrations.

The analytic estimates also agree with Bonsor18 at the level of the direct physical collision probability. If post-resonance diffusion process is not included, our collision fraction has the same parameter dependence as their rate-based estimate, as shown in Equation~\eqref{eq:f_col_p}. However, this estimate underpredicts the collision fractions measured in the numerical simulations. This difference highlights the importance of the post-resonance stochastic phase: after resonance escape, diffusion in orbital energy can increase the cumulative probability of planetary collision before the particle is removed by sublimation or ejection.

Our model captures this process by deriving the encounter statistics directly from the impact parameter distribution and the corresponding kick model, $\Delta x(b,\chi)$, and using them to construct a Monte Carlo description of orbital diffusion. In contrast, Bonsor18 describe accretion and ejection as independent Poisson processes with approximately constant rates over an interaction interval. In our framework, no explicit ejection probability per orbit is introduced; instead, all non-collisional outcomes emerge self-consistently from the diffusion process. This stochastic diffusion following resonance escape enhances the collision probability and is therefore essential for reproducing the accretion fractions observed in the numerical simulations.

\section{Summary}\label{sec7}
We have studied the dynamical evolution and ultimate fates of dust particles
migrating inward through close-in star--planet systems under the combined
effects of radiation pressure and PR drag. The primary motivation of this study is to compute the efficiency with which dust inpiraling from distant reservoirs may be accreted into close-in planetary upper atmospheres. Our main
results are as follows.

\begin{enumerate}

\item Capture into the dissipative equilibrium of mean motion resonances is investigated through orbit integrations over a range of planet mass ($m_1$) and radiation parameter ($\beta$). Over the range studied, it is found that capture into  $2\!:\!1$ and $3\!:\!2$  occurs most frequently. Capture is suppressed for sufficiently large planet-to-star mass ratio $\ga 10^{-2}$ for all values of $\beta$.

\item Resonant evolution including PR drag requires a treatment accurate at high eccentricity. We carried out a numerical average of the disturbing function over the rapid phase variable, and with no small eccentricity expansion. This was used in Lagrange’s planetary equations to determine equilibrium points, as well as the libration frequencies and growth rates for small perturbations. These equilibrium points are compared to the exact orbital integrations, and are found to describe the evolution well for slow migration rates and away from bifurcations. The failure of low order expansions is discussed in an appendix.

\item Dissipative equilibrium points are generically overstable for dust particles under PR drag. 
We linearized the dissipative
equations of motion about the equilibrium point,  and found  growing modes for a range of important resonances (including $3\!:\!1$, $2\!:\!1$, $3\!:\!2$, etc). We also derived a simplified analytic expression for the growth rate that agrees well with the numerical linearization results. As a consequence, resonant capture is temporary, and dust particles ultimately escape resonance over broad ranges of $m_1$ and $\beta$.

\item Dust particles exit resonance on planet-crossing orbits with encounters at effectively random phases. Hence post-resonance evolution is stochastic and may be modeled statistically. Orbit integrations are used to determine the distribution  of impact parameters $P(b)$ for close encounters, and a toy model is used to explain the results. We find that, at small $b$, $P(b)$ is constant for nearly coplanar orbits, and $P(b) \propto b$ for inclined orbits. The distribution of orbital energy changes $P(\Delta x)$ can be explained from $P(b)$, and shows power-law tails  $|\Delta x|^{-2}$ (coplanar) and $|\Delta x|^{-3}$ (inclined).

\item The distribution of orbital energy changes for both close and distant encounters is incorporated into a Monte Carlo simulation for post-resonance evolution, including the effect of PR drag, physical collisions with the planet, sublimation near the star, and ejection from the system.

For ensembles of $N=10^4$ dust particles, the resulting final fate fractions are computed as a function of planetary orbital separation and show good agreement with full orbit integrations. 

\item Monte Carlo results are presented for two planetary systems, a hot Jupiter around a Sun-like star with a coplanar dust orbits, and a low mass star and small planet with inclined dust. In both cases, accretion efficiency onto the planet may be quite high near the star, 10-90\%, for the range of orbital separation used. For planets too near the star, dust on highly eccentric orbits may be predominantly sublimated near the star. For massive planets further from the star, ejections become important.

\end{enumerate}

Overall, our results provide a physical framework for dust migration through mean motion resonances and the subsequent stochastic scattering phase. By deriving encounter statistics from first principles and incorporating them into a Monte Carlo framework, we predict collision, sublimation, and ejection fractions in good agreement with $N$-body simulations. This approach connects the final dust fates to the underlying resonant dynamics, close-encounter geometry, and PR drag evolution, yielding improved constraints on dust fate. These results provide a basis for interpreting possible dust-driven atmospheric enrichment in close-in exoplanetary systems.

\software{SymPy \citep{Meurer2017SymPy},
          SciPy \citep{2020SciPy-NMeth},
          REBOUND \citep{2012A&A...537A.128R},
          REBOUNDx \citep{2020MNRAS.491.2885T}
          }

\begin{acknowledgments}
We thank Yifan Zhou, Shane Davis, Zhi-Yun Li, Chase Funkhouser, Eonho Chang, Dominic Samra, and Peter Gao for useful discussions. This work used High-Performance Computing systems Rivanna and Afton at the University of Virginia. This research was supported by NASA ATP grant 80NSSC18K0696, ``Exoplanetary MHD Outflows Driven by EUV Heating, Lyman alpha Radiation Forces and Stellar Tides.” We acknowledge funding from the Virginia Institute for Theoretical Astrophysics (VITA), supported by the College and Graduate School of Arts and Sciences at the University of Virginia.
\end{acknowledgments}

\vspace{5mm}

\appendix

\section{ Low Order Expansions }\label{app:loworderexpansion}

Previous investigations (e.g., \citealt{1993Icar..104..244W}; \citealt{2015MNRAS.448..684S}) have used low order in $e_2$ approximations to try to understand the dynamics. This is useful as the particle enters resonance and $e_2 \sim (m_1/m_0)^{1/3} \ll 1$. However, the dissipative equilibrium for PR drag has $e_2 \sim {\cal O}(1)$, and as we will show, low order expansions produce qualitatively and quantitatively incorrect results: the on-axis equilibrium points initially followed by the system become unstable; new off-axis equilibrium points arising from higher harmonics in the disturbing function appear and are stable; and the stability of the dissipative equilibrium points is incorrect using the low-order expansions.

\subsection{First-Order Approximation}\label{app:firstorder}

In this section we will discuss an approximation in which LPE, $R_2$, and the PR terms use leading order in $e_2$ expansions (``${\cal O}(e_2)$ LPE"). For a particle in the vicinity of a $j\!:\!j-1$ exterior mean motion resonance, the disturbing function at ${\cal O}(e_2)$ is given by
\begin{equation}
R_2 = \mu_1a_2^2n_2^2C_1e_2\cos{\phi_2},
\label{eq:R2_1st_order}
\end{equation}
where $\mu_1 = m_1/[m_0(1-\beta)]$, 
\begin{equation}
C_1 = \frac{1}{2}[2j-1+\alpha D]b_{1/2}^{(j-1)}-\frac{1}{2\alpha^2}\delta_{j,2}
\end{equation}
is the function of Laplace coefficients evaluated at $\alpha = a_1/a_{\rm 2, res}$, 
$b_{1/2}^{(1)}(\alpha)$ denotes a Laplace coefficient, and $D=d/d\alpha$.
Here $C_1 \simeq 0.428$ for $\beta=0$ and $j=2$.

Expanding Equations \eqref{eq:dadt}, \eqref{eq:dedt}, and \eqref{eq:dphidt} to leading order in $e_2$, and plugging in Equation \eqref{eq:R2_1st_order}  gives the simplified equations of motion
\begin{equation}\label{eq:dndt1st}
\frac{d n_2}{dt} = 3j \mu_1 n_2^2 C_1e_2\sin{\phi_2} + \frac{n_2}{\tau_{n_2}}(1+3e_2^2),
\end{equation}
\begin{equation}\label{eq:dedt1st}
\frac{de_2}{dt} = -
\mu_1n_2C_1\sin{\phi_2} - \frac{e_2}{\tau_{e_2}},
\end{equation}
\begin{equation}\label{eq:dphidt1st}
\frac{d\phi_2}{dt} = jn_2 - (j-1)n_1 - \frac{\mu_1 n_2}{e_2}C_1\cos{\phi_2}.
\end{equation}
For the moment, we allow $\tau_{e2}/\tau_{n2}$ to be an arbitrary ratio, rather than PR value $6/5$.
Here all factors of $n_2$ have been evaluated at $n_{\rm 2, res}$ except for the $dn_2/dt$ in Equation \eqref{eq:dndt1st} and the first term on the right hand side of Equation \eqref{eq:dphidt1st}. This could be made more explicit if $\kappa_2$ in Equation \eqref{eq:kappa2} was used as a variable instead of $n_2$. At this order, $a_2/a_{\rm 2, res} \simeq 1 + \kappa_2 + je_2^2$ and so to leading order $n_2 \simeq n_{\rm 2, res}$, and we suppress the subscript for simplicity.

The last term $(3e_2^2n_e/\tau_{n2})$ in Equation \eqref{eq:dndt1st} requires explanation. At first this seems to be a higher order term that can be neglected in that equation. However, when the coefficients are evaluated at the equilibrium points, it is comparable to the first term and makes an order unity contribution to the equilibrium eccentricity. Inclusion of this term also agrees with the straightforward expansion of Equation \eqref{eq:kappa2dot}. This term was shown to be the key to overstability for the exterior perturber case studied in \citet{2014AJ....147...32G}.

First, we find the conservative equilibrium points by ignoring PR terms and setting $de_2/dt=0$ in Equation \eqref{eq:dedt1st} and $d\phi_2/dt=0$ in Equation \eqref{eq:dphidt1st}. We also replace $jn_2-(j-1)n_1$ with $(-3jn_2/2)(\kappa_2 + je_2^2)$. The ``on-axis" conservative equilibrium points have $\phi_{\rm 2, eq}=0,\pi$, with eccentricity determined by the cubic equation
\begin{equation}\label{eq:cubic_for_e2}
3j^2 e_2^3 + 3j\kappa_2 e_2 + 2\mu_1 C_1 \cos \phi_{\rm 2, eq} =  0.
\end{equation}
As the particle enters resonance at large and positive $\kappa_2$, there is one equilibrium point with $e_2(\kappa_2) \simeq (2\mu_1 C_1)/(3j\kappa_2) \ll 1$. The ``on-axis bifurcation" occurs when the discriminant of the cubic is zero at $\kappa_{\rm 2, on-axis} = - (3\mu_1^2C_1^2/j)^{1/3}$, at which point $e_2(\kappa_{\rm 2, on-axis}) = (8\mu_1 C_1/(3j^2))^{1/3}$. Two new equilibrium points appear at the bifurcation with $\phi_2=0$ and $e_2(\kappa_{\rm 2, on-axis}) = (\mu_1 C_1/(3j^2))^{1/3}$. Well after the bifurcation, when $\kappa_2$ is large and negative, the solution at $\phi_2=\pi$ has $e_2(\kappa_2) \simeq (|\kappa_2|/j)^{1/2}$ while the two solutions at $\phi_2=0$ have large  $e_2(\kappa_2) \simeq (|\kappa_2|/j)^{1/2}$ and small $e_2(\kappa_2) \simeq (2\mu_1 C_1)/(3j|\kappa_2|)$.

Treating $\kappa_2$ as a fixed parameter and perturbing about the conservative equilibrium points gives the libration frequency
\begin{eqnarray}\label{eq:omega_0_1st}
\omega_0^2 & = & \left( \frac{\mu_1 C_1 n_2}{e_2} \right)^2 - 3j^2 \mu_1 C_1 n_2^2 e_2 \cos \phi_2.
\end{eqnarray}
The root with $\phi_2=\pi$ is always stable and has minimum libration frequency just before the bifurcation with $\omega_{\rm 0, \min} = (3^{5/6}/2^{1/3})(j\mu_1 C_1)^{2/3}n_2$. This behavior underlies the adiabatic capture criterion used by \citet{2014AJ....147...32G}. In our problem, however, this estimate is not applicable: the libration frequency does not attain a finite minimum, but instead passes through zero when the off-axis bifurcation occurs (see section \ref{sec5}). For the roots with $\phi_2=0$, their libration frequency is zero at the bifurcation. The small $e_2$ root is stable with $\omega_0$ increasing away from the bifurcation, whereas the large $e_2$ is always unstable with $\omega_0^2 < 0$.

Next, we find the dissipative equilibrium by including the PR terms. Setting $d n_2/dt=0$ and $de_2/dt = 0$ in Equations \eqref{eq:dndt1st} and \eqref{eq:dedt1st}, we obtain the equilibrium eccentricity and resonant angle
\begin{equation}\label{eq:e2_eq_1st}
e_{2, \mathrm{eq}}  = 
{\left(\frac{\tau_{e_2}/\tau_{n_2}}{3j-\frac{3\tau_{e_2}}{\tau_{n_2}}}\right)}^{1/2} =  \left(\frac{2}{5j-6}\right)^{1/2}
\end{equation}
and
\begin{equation}\label{eq:phi2_eq_1st}
\sin{\phi}_{2, \mathrm{eq}} = -\frac{e_{\rm 2, eq}}{\mu_1 C_1 n_2 \tau_{e2}}.
\end{equation}
Notice that for the $2\!:\!1$ resonance, $e_{\rm 2, eq}=2^{-1/2}\simeq 0.7$, while \citet{1993Icar..104..244W} found the incorrect value $e_{\rm 2, eq}=5^{-1/2}\simeq 0.45$, which happens to be closer to the correct value. This dissipative equilibrium point is shifted away from the conservative equilibrium point $\phi_2=\pi$ due to the eccentricity damping term. The corresponding value of $\kappa_{\rm 2, eq}$ is then determined by $e_{\rm 2, eq}$ through Equation \eqref{eq:kappa2}. In order for the solution in Equation \eqref{eq:phi2_eq_1st} to be self-consistent, the right hand side must have magnitude less than 1, which places an upper limit on the damping rate  $\tau_{e2}^{-1} \lesssim (\mu_1 C_1 n_2)/e_{\rm 2, eq}$.

Linearizing Equations \eqref{eq:dndt1st}, \eqref{eq:dedt1st}, and \eqref{eq:dphidt1st} around the equilibrium values $n_{\rm 2, eq}$, $e_{2,\mathrm{eq}}$ and $\phi_{2,\mathrm{eq}}$, the perturbations $(\delta n_2, \delta e_2, \delta \phi_2)$ satisfy a 3x3 linear system, for which there are three eigenmodes. We find that all three eigenmodes are damped, implying that the dissipative equilibrium point is stable. This result is incorrect, as demonstrated numerically in Section 
\ref{sec4} and analytically below. In detail, we find a complex conjugate pair with $\omega = \pm \omega_0$ and growth rate $\gamma \simeq -3/(2t_{\rm pr})$, together with a non-oscillatory mode with zero libration frequency ($\omega=0$) and growth rate $\gamma \simeq -2/t_{\rm pr}$. These approximate damping rates assume $e_{\rm 2, eq} \gg (\mu_1 C_1)^{1/3}$ at the dissipative equilibrium.

That the equilibrium points at this order are always stable also contrasts with the result of \citet{2014AJ....147...32G}, who consider an outer perturber. In their Equation 29, the two terms in the growth/damping rate have opposite signs, so the overall sign depends on the competition between $\tau_{e}/\tau_{n}$ and the planet mass. However, for an inner perturber the two terms have the same sign.

\citet{1993Icar..104..244W} (hereafter WJ93) also employed the ${\cal O}(e_2)$  planetary equations, but used the exact PR drag terms. They first derived $\sin \phi_{\rm 2, eq}$ by setting $\dot{a}_2=0$, and then plugged this result into the $\dot{e}_2$ equation. They then took the small $e_2$ limit to derive the simple solution
\begin{equation}\label{eq:e2_t}
e_2(t) = \sqrt{\frac{2}{5j}\left[1-\exp(-\frac{t-t_0}{\tau})\right]},
\end{equation}
where $t_0$ is the time of resonant capture, and the time constant $\tau=a_1^2c(\beta/(1-\beta)^{2/3})((j-1)/j)^{4/3}/(5Gm_0)$. Although this equation gives the correct qualitative behavior, the equilibrium point $(e_{\rm 2, eq},\phi_{\rm 2, eq})=(\sqrt{2/(5j)},\pi)$ is incorrect. The discrepancy arises from using low order planetary equations, irrespective of the approximation for the disturbing function. However, the small $e_2$ expansion is not necessary. In Equations \eqref{eq:dadt} and \eqref{eq:dedt}, the derivatives of the $Q$-averaged $\langle R_2 \rangle$ are related as $R_2$ only depends on the combination in $\phi_2$. Hence Equation \eqref{eq:dadt} can be solved for $\partial R_2/\partial \phi_2$, and this result can be used to eliminate the $R_2$ derivatives in Equation \eqref{eq:dedt}. This gives an equation for $de_2/dt$ that is independent of the explicit form of $R_2$ and has the correct equilibrium point. Expanding that expression for small $e_2$ places the equilibrium point for the $2\!:\!1$ resonance at $e_{\rm 2, eq}=2^{-1/2}$, as derived above.

\begin{figure*}[ht!]
    \centering
    \includegraphics[width=\textwidth]{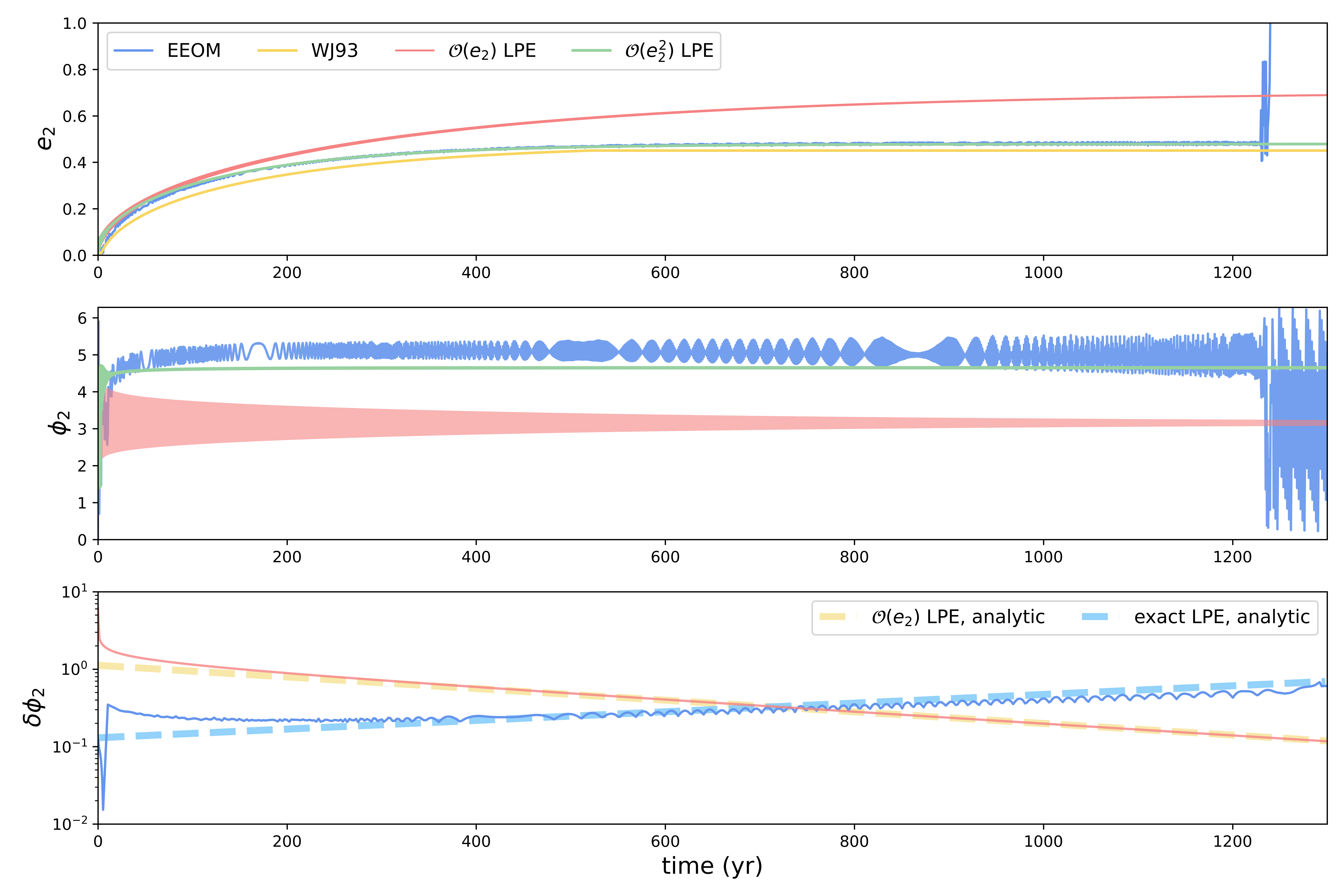}
    \caption{Evolution of eccentricity $e_2$ (top panel), resonant angle $\phi_2$ (middle panel), and 
    $\delta \phi_2 = \phi_2 - \phi_{\rm 2, eq}$ (bottom panel) for a particle in $2\!:\!1$ MMR ($\beta=0.01$). The EEOM numerical integration (blue curves) is compared with several different approximations: the WJ93 solution (yellow curve in top panel; Equation~\ref{eq:e2_t}); the ${\cal O}(e_2)$ LPE (red curves; Equations~\ref{eq:dndt1st}–\ref{eq:dphidt1st}); and the hybrid ${\cal O}(e_2^2)$ LPE (green curves; Appendix \ref{app:secondorder}). 
    The bottom panel shows $\delta \phi_2$ for the EEOM integration (thin solid blue line) and ${\cal O}(e_2)$ LPE integration (thin solid red line), compared to the growth/decay slopes calculated  from the linearized equations:
    ${\cal O}(e_2)$ LPE (yellow dashed line;  $\delta \phi_2 \propto e^{-3t/2t_{\rm pr}}$);
    exact LPE and $Q-$averaged $\langle R_2 \rangle$ (blue dashed line; $e^{\gamma t}$ with $\gamma$ from Equation~\ref{eq:growthrate_analytic}).
    }
    \label{fig:odeplot}
\end{figure*}

Figure~\ref{fig:odeplot} shows solutions for $e_2(t)$ (top panel), $\phi_2(t)$ (middle panel), and $\delta \phi_2 = \phi_2 - \phi_{\rm 2, eq}$ (bottom panel) versus time near the $2\!:\!1$ MMR. Several different solution methods are compared: ${\cal O}(e_2)$ LPE (red curves; Eqs.~\ref{eq:dndt1st}–\ref{eq:dphidt1st}); the WJ93 solution (yellow curve in top panel; Equation~\ref{eq:e2_t}), and  EEOM numerical integration (blue curves; Section \ref{sec2.1}). Equation \eqref{eq:e2_t} shows qualitative and rough quantitative agreement with the exact result. The correct ${\cal O}(e_2)$ LPE solution has both $e_{\rm 2, eq}$ and $\phi_{\rm 2, eq}$ significantly in error.
These discrepancies indicate that first-order disturbing-function solutions lack sufficient accuracy, necessitating the inclusion of higher-order terms in both $R_2$ and the LPE.

\subsection{Second- and Higher-Order Disturbing Functions}\label{app:secondorder}

Motivated by the appearance of off-axis equilibrium points, here we consider a hybrid higher order expansion, called ``hybrid ${\cal O}(e_2^2)$ LPE", in which $R_2$ uses resonant terms up to ${\cal O}(e_2^2)$, but the exact expressions are used for LPE and also for the PR terms. As we will show, including the second harmonic term in $R_2$ gives rise to the off-axis equilibrium points which the system follows for $e_2 \ga 0.03$. The use of the exact LPE and PR drag terms gives the correct $e_{\rm 2, eq}$, even though only a low-order approximation to $R_2$ is used.

We focus again on the $2\!:\!1$ resonance ($j=2, k=1$), now including the second-harmonic term
\begin{equation}\label{eq:R2_2nd}
R_2 = \mu_1a_2^2n_2^2(C_1e_2\cos{\phi_2}+C_2e_2^2\cos{2\phi_2}),
\end{equation}
where $C_2 = (1/8)[38+14\alpha D + \alpha^2 D^2]b_{1/2}^{(2)} \simeq 3.594$ for $\beta=0$. 
Equation \eqref{eq:R2_2nd} is inserted into Equations \eqref{eq:dadt}, \eqref{eq:dedt}, and \eqref{eq:dphidt}. The $\partial R_2/\partial a_2$ derivative requires derivatives of $C_1$ and $C_2$, not shown here for brevity.

The key new feature, due to the second harmonic term in $R_2$,  is the appearance of the off-axis equilibrium points. Ignoring PR drag, the conservative $e_{\rm 2, eq}$ is found by setting $da_2/dt=0$ in Equation \eqref{eq:dadt} and $de_2/dt=0$ in Equation \eqref{eq:dedt}, which both give the condition
\begin{eqnarray}
\sin{\phi_2} \left( C_1+ 4C_2e_2 \cos{\phi_2} \right) & = & 0.
\end{eqnarray}
The first factor again gives the on-axis equilibrium point at $\phi_{\rm 2, eq}=\pi$, and the two additional points at  $\phi_{\rm 2, eq}=0$ after the on-axis bifurcation. But when $e_2 \geq e_{\rm 2, cr} \equiv C_1/(4C_2) \simeq 0.03$, two new off-axis equilibrium points appear at $\phi_{\rm 2, eq}=\pi$, and move to the right as $\cos(\phi_{\rm 2, eq})=-e_{\rm 2, cr}/e_2$, as seen in Figures \ref{fig:phasespace1} and \ref{fig:phasespace2}. We find that the second harmonic term in $R_2$ captures the evolution near the bifurcation well, consistent with the findings of \cite{2005ApJ...619..623M}.
However, that the approximate off-axis points stay in the second and third quadrants is an inadequacy of the low-order $R_2$ used. The exact result shows that these equilibrium points eventually move to the first and fourth quadrants as $\kappa_2$ decreases. As discussed above, Equation \eqref{eq:dadt} can be solved for $\partial R_2/\partial \phi_2$, which can then be substituted into Equation \eqref{eq:dedt}. This eliminates the explicit dependence on the $Q$-averaged $\langle R_2 \rangle$ and gives the correct $e_{\rm 2, eq}$ for the dissipative equilibrium, as long as the LPE and PR drag expressions are exact.

The solutions for the conservative $e_{\rm 2, eq}$ are found by plugging Equation \eqref{eq:R2_2nd} into Equation \eqref{eq:dphidt}. Although there is no analytic solution for large $e_2$, including the second harmonic term in Equation \eqref{eq:cubic_for_e2} and setting $j=2$ gives
\begin{equation}
e_2^3 + \frac{1}{2} \left( \kappa_2 + \frac{\mu_1 C_1 \cos(2\phi_2)}{6 e_{\rm 2, cr} } \right) e_2 + \frac{1}{6} \mu_1 C_1 \cos (\phi_2) = 0.
\end{equation}
The equilibrium eccentricities are then obtained by evaluating this equation at equilibrium values of $\phi_{\rm 2, eq}$ and solving for $e_{\rm 2, eq}$.

For the on-axis roots, $\cos(\phi_{\rm 2, eq})=\pm 1$ and $\cos(2\phi_{\rm 2,eq})=1$. There are again one or three 
solutions for the conservative equilibrium points, depending on $\kappa_2$ versus $\kappa_{\rm 2, on-axis}$.  Setting the discriminant to zero gives the slightly modified on-axis bifurcation point
\begin{eqnarray}
    \kappa_{\rm 2, on-axis} & = & - \left( \frac{3\mu_1^2C_1^2}{2} \right)^{1/3} - \frac{\mu_1 C_1}{6e_{\rm 2, cr}}.
    \label{eq:kap_on_axis_2nd}
\end{eqnarray}
The values of $e_2(\kappa_2)$ for each solution are slightly shifted due to the second harmonic term.

For the off-axis roots,  we  plug in $\cos(\phi_{\rm 2, eq})=-e_{\rm 2, cr}/e_2$ and $\cos(2\phi_{\rm 2,eq})=2(e_{\rm 2, cr}/e_2)^2-1$ to find a quadratic equation for $e_2$ with solution
\begin{eqnarray}
e_{\rm 2, eq}(\kappa_2) & = & \sqrt{ \frac{\mu_1 C_1}{12 e_{\rm 2, cr}} - \frac{\kappa_2}{2} }
\end{eqnarray}
Setting $e_{\rm 2, eq}=e_{\rm 2, cr}$, the off-axis bifurcation occurs at
\begin{equation}\label{eq:kappa2_offaxis}
\kappa_{2,\mathrm{off\mbox{-}axis}} = \frac{\mu_1C_1}{6e_{2, \mathrm{cr}}} - 2e_{2, \mathrm{cr}}^2.
\end{equation}
The dependence on $\mu_1$ is different between Equations \eqref{eq:kap_on_axis_2nd} and \eqref{eq:kappa2_offaxis}. Equating these two expressions, we find a critical planet mass ratio
\begin{eqnarray}
    \mu_{\rm 1, cr} & = & \frac{3e_{\rm 2, cr}^3}{2C_1} \simeq 10^{-4}.
\end{eqnarray}
For $\mu_1 \geq \mu_{\rm 1, cr}$, the off-axis bifurcation occurs first during the evolution, while the on-axis bifurcation occurs first for $\mu_1 \leq \mu_{\rm 1, cr}$. This explains the trends shown in Figures \ref{fig:phasespace1} and \ref{fig:phasespace2}.

The libration frequency and growth rate can be derived analytically keeping lowest order terms in $e_2$. Treating $\kappa_2$ as a constant and perturbing around the conservative off-axis equilibrium points leads to an approximate libration frequency
\begin{equation}\label{eq:omega_approx}
\omega_0^2 \simeq 12 \mu_1 C_1 n_2^2e_{\rm 2, cr}(e_2^2/e_{\rm 2, cr}^2 - 1),
\end{equation}
which is zero at the off-axis bifurcation and grows thereafter. Perturbing about the dissipative equilibrium point $e_{\rm 2, eq}=2^{-1/2}$ and including PR drag as a small perturbation leads to the same (incorrect) damping rate $\gamma=-3/(2t_{\rm pr})$ as for the ${\cal O}(e_2)$ LPE. 

The results obtained by integrating the hybrid ${\cal O}(e_2^2)$ LPE 
are shown in Figure~\ref{fig:odeplot} (green curves). The eccentricity evolution agrees well with EEOM integration, as the equilibrium eccentricity is  determined by the LPE and PR terms and is independent of the specific form of $R_2$ used.
The resonant angle $\phi_2$ reaches $\phi_{\rm 2,eq} \simeq 3\pi/2$, 
as expected for $e_2 \gg e_{\rm 2, cr}$, notably different than the EEOM result. Moreover, the libration angle $\delta\phi_2$ is damped as for the ${\cal O}(e_2)$ LPE, once again contradicting the numerical EEOM results. 

These discrepancies prompt us to include even higher-order harmonic terms in the disturbing function
\begin{equation}
R_2 = \frac{Gm_1}{a_2} \sum_{n=1}^{n_{\rm max}} C_n e_2^n \cos{(n\phi_2)}.
\end{equation}
Through numerical experimentation, we find that retaining at least the first four harmonics ($n_{\rm max} = 4$) is required to reproduce overstable librations, i.e., a growing $\delta\phi_2$, in the analytical ODE framework.  
This result demonstrates that higher-order terms in the disturbing function are essential to get the correct $\phi_{\rm 2, eq}$ and overstability.

\section{Derivation of the Impact Parameter Distribution}
\label{app:impact_parameter}

In the rotating frame, the planet is fixed at $\mathbf{P}=(a_1,0,0)$ in the star--planet plane. We consider the general case of nonzero dust inclination. The test particle orbit lies in a plane obtained from the $xy$–plane by the rotation $R_z(\Omega_2)R_x(i_2)$. Its unit normal is therefore
\begin{equation}
\mathbf{n}
=R_z(\Omega_2)R_x(i_2)\hat{\mathbf z}
=\bigl(\sin i_2 \sin\Omega_2,\ -\sin i_2 \cos\Omega_2,\ \cos i_2\bigr).
\end{equation}
The guiding center is a circle of radius $a_2$ in this tilted dust plane:
\begin{equation}
\mathbf{C}(\psi)=R_z(\Omega_2)R_x(i_2)(a_2\cos\psi,\ a_2\sin\psi,\ 0),
\end{equation}
where the guiding-center phase $\psi$ is taken to be uniformly distributed on $[0,2\pi)$. At fixed $\Omega_2$, the perpendicular offset of the planet from the dust plane is
\begin{equation}
\delta = |(\mathbf{P}-\mathbf{C})\cdot \mathbf{n}| = a_1 |\sin i_2 \sin\Omega_2|,
\end{equation}
which is independent of $\psi$ because $\mathbf C(\psi)\cdot\mathbf n=0$. Let $\mathbf{P_\parallel}$ denote the orthogonal projection of $\mathbf{P}$ onto the dust plane. Its in-plane radius is
\begin{equation}
A \equiv |\mathbf{P_\parallel}|=\sqrt{a_1^2-\delta^2}.
\end{equation}
The in-plane separation between $\mathbf{P_\parallel}$ and the guiding center is then
\begin{equation}
r(\psi)=\sqrt{a_2^2+A^2-2a_2A\cos(\psi-\psi_0)},
\end{equation}
where $\psi_0$ is the phase at which the guiding center is closest to $\mathbf{P_\parallel}$. Since $\psi$ is uniformly distributed, the phase shift $\psi-\psi_0$ does not affect the statistics. The distribution of $r(\psi)$ is then
\begin{equation}
p_r(r)
=\frac{2r}{\pi\sqrt{\bigl((a_2+A)^2-r^2\bigr)\bigl(r^2-(a_2-A)^2\bigr)}},
\end{equation}
where $r_{\min} = |a_2-A|$, $r_{\max} = |a_2+A|$.
At fixed $\delta$ (hence fixed $\Omega_2$), the closest approach between the planet and the tilted epicycle is
\begin{equation}
b=\sqrt{\delta^2+(r-\rho)^2} .
\end{equation}
Let $\eta=|r-\rho|$, the corresponding PDF for a fixed $\Omega_2$ is therefore
\begin{equation}
p^{(\Omega_2)}(b)
=\frac{b}{\sqrt{b^2-\delta^2}}
\Bigl[
p_r\bigl(\rho+\eta\bigr)
+p_r\bigl(\rho-\eta\bigr)
\Bigr].
\end{equation}
The second term only exists when the projected planet is accessible by the guiding-center circle, e.g., $\rho-\sqrt{b^2-\delta^2}\in(r_{\min}, r_{\max})$. To obtain the full inclined distribution, we average over $\Omega_2$. With $i_2$ fixed,
\begin{equation}
\delta = a_1 |\sin i_2 \sin\Omega_2|\in[0,\delta_{\max}],\qquad \delta_{\max}=a_1\sin i_2,
\end{equation}
For $\Omega_2$ uniformly distributed on $[0,2\pi)$, $\delta$ has the arcsine density on $[0, \delta_{\max}]$:
\begin{equation}
p_\delta(\delta)
=\frac{2}{\pi}\frac{1}{\delta_{\max}\sqrt{1-(\delta/\delta_{\max})^2}}.
\end{equation}
The $\Omega_2$-averaged impact parameter distribution is then
\begin{equation}\label{eq:p_b_simp}
P_{i_2\neq0}(b)
=\int_0^{\delta_{\max}} p^{(\Omega_2)}(b)\, p_\delta(\delta)\, d\delta.
\end{equation}
A convenient substitution is $\delta=b\sin\theta$, so that $\theta_{\max}=\arcsin\bigl(\min\{1,\ \delta_{\max}/b\}\bigr)$. The distribution becomes
\begin{equation}
p_\delta(\delta) d\delta = \frac{2}{\pi \delta_{\max}} 
\frac{b d\theta}{\sqrt{1-\frac{b^2}{\delta_{\max}^2}\sin^2\theta}}.
\end{equation}
With this substitution, Equation~\eqref{eq:p_b_simp} reduces to a single 1-D integral:
\begin{equation} \label{eq:P_of_b_inclined}
P_{i_2\neq0}(b)
=\frac{2b}{\pi \delta_{\max}}
\int_{0}^{\theta_{\max}}
\frac{ \displaystyle
p_r\bigl(\rho+b\cos\theta\bigr)
+
p_r\bigl(\rho-b\cos\theta\bigr)
}
{\sqrt{1-\frac{b^2}{\delta_{\max}^2}\sin^2\theta}}\ d\theta.
\end{equation}
This equation is used in Figure \ref{fig:P_b_P_dx} for the analytic lines of $P(b)$ in the top panels. If $\rho$ lies strictly inside the planar annulus $|a-a_1|<\rho<a+a_1$, then for $b\to 0$
\begin{equation}\label{eq:P_b_inc}
\begin{split}
P_{i_2\neq0}(b)
&=\frac{2}{\delta_{\max}}\, p_r(\rho)\, b
\;+\;\mathcal{O}(b^3)\\
&\simeq\frac{b}{a_1\sin{i_2}}\frac{4\rho}{\pi\sqrt{\bigl((a_2+a_1)^2-\rho^2\bigr)\bigl(\rho^2-(a_2-a_1)^2\bigr)}}\\
&=\frac{b}{a_1\sin{i_2}}C_0.
\end{split}
\end{equation}
Specifically, in the $\delta_{\max} \to 0$ limit, Equation \eqref{eq:P_of_b_inclined} collapses to the coplanar ($i_2=0$) expression
\begin{equation}
\begin{aligned}
& \quad \quad P_{i_2=0}(b)\\
&= \frac{2(b+\rho)}{\pi
\sqrt{\bigl((a_2+a_1)^2-(b+\rho)^2\bigr)
       \bigl((b+\rho)^2-(a_2-a_1)^2\bigr)}} \\
&\quad + \frac{2(\rho-b)}{\pi
\sqrt{\bigl((a_2+a_1)^2-(\rho-b)^2\bigr)
       \bigl((\rho-b)^2-(a_2-a_1)^2\bigr)}}.
\end{aligned}
\end{equation}
The second term only exists if the epicycle can reach the planet’s circle of radius, i.e., for $0\le b \le \rho-|a_2-a_1|$. In our model, we restrict attention to this regime, where both branches contribute and the distribution exhibits a characteristic small–$b$ plateau $C_0$:
\begin{equation}
P_{i_2=0}(b=0)
= \frac{4\rho}{\pi\sqrt{\big((a_2+a_1)^2-\rho^2\big)\big(\rho^2-(a_2-a_1)^2\big)}}.
\end{equation}
Thus, whereas the coplanar case exhibits a constant plateau near $b=0$, a nonzero inclination smooths this plateau into a linear rise $P_{i_2\neq0}(b)\propto b$, after averaging over $\Omega_2$. The small-$b$ approximation for both coplanar and inclined cases can be summarized in the unified expression:
\begin{equation}
P_{b \rightarrow 0}(b) = C_0\frac{b}{\sqrt{b^2 + (a_1\sin{i_2})^2}}.
\end{equation}

\section{Derivation of the Energy Change Distribution}
\label{app:pdx}

In the rotating frame, the dimensionless Jacobi integral is
\begin{equation}
C_J = -\frac{2}{v_1^2}\left[\frac{1}{2}u^2-\frac{Gm_0(1-\beta)}{r_{02}}-\frac{Gm_1}{r_{12}}-\frac{1}{2}n_1^2(x_2^2+y_2^2)\right],
\end{equation}
where $u=|\mathbf{u}|=|\mathbf{v_2}-\mathbf{v_1}|$ is the relative velocity between the dust particle and the planet, $x_2$ and $y_2$ are the particle's Cartesian coordinates. Taking the $m_1 \ll m_0$ limit and evaluating near $x_2 \simeq a_1$, $y_2=0$, we find
\begin{equation}
C_J = -\frac{u^2}{v_1^2}+2(1-\beta)+1   
\end{equation}
gives the ``relative velocity at infinity" of the dust particle and planet in terms of $C_J$ to be \citep{1976iecg.book.....O}
\begin{equation}
u_{\infty} = v_1 \sqrt{3 - 2\beta - C_J}.
\end{equation}
The orbital energy of the dust particle relative to the star (neglecting the planet) is
\begin{equation}
\varepsilon = -\frac{G m_0 (1-\beta)}{2 a_2}
\equiv -\frac{1}{2}(1-\beta)\,v_1^2\,x.
\end{equation}
During an encounter with impact parameter $b$, the planet imparts a velocity kick \citep{2008gady.book.....B}
\begin{equation}\label{eq:delta_u}
\Delta \mathbf{u}
= -2\,\frac{\mathbf{u} + u\,\mathbf{b}/b_0}
{1+(b/b_0)^2},
\end{equation}
at $\mathbf{u} \dotproduct \mathbf{b} = 0$, where the impact parameter vector is $\mathbf b = |\mathbf{x}_2-\mathbf{x}_1|_{\rm min}$, and the impact parameter for 90$^\circ$ scattering is $b_0 = G m_1 / u^2$.  
While the encounter conserves the energy of the dust particle relative to the planet, it modifies the particle’s orbital energy with respect to the star. Approximating the dust particle as stationary during the encounter, the energy change of the dust particle around the star is
\begin{equation}
\begin{aligned}
\Delta \varepsilon
&= -\frac{1}{2}(1-\beta)\,v_1^2\,\Delta x \nonumber\\
&= \frac{1}{2}\left(\mathbf{v}_1 + \mathbf{u} + \Delta \mathbf{u}\right)^2
   - \frac{1}{2}\left(\mathbf{v}_1 + \mathbf{u}\right)^2 \nonumber\\
&= \mathbf{v}_1 \cdot \Delta \mathbf{u}.
\end{aligned}
\end{equation}
Using Equation~\eqref{eq:delta_u}, the corresponding change in $x$ is therefore
\begin{align}
\Delta x
&= -\frac{2}{(1-\beta)\,v_1^2}\,
\mathbf{v}_1 \cdot \Delta \mathbf{u} \nonumber\\
&= \frac{4}{(1-\beta)\left[1+(b/b_0)^2\right]v_1^2}
\left(
\mathbf{v}_1 \cdot \mathbf{u}
+ \frac{u}{b_0}\,\mathbf{v}_1 \cdot \mathbf{b}
\right).
\end{align}
The dominant contribution arises from the component of the kick perpendicular to the incoming relative velocity (the second term in the parenthesis). Writing $\mathbf{v}_1 \cdot \mathbf{b}
= v_1 b \cos\chi$, where $\chi$ is the encounter phase angle, and assuming $\chi$ is uniformly distributed, $P(\chi)\,d\chi = d\chi/(2\pi)$, we obtain
\begin{equation}
\Delta x(b,\chi)
= \Delta x_0\,\frac{2(b/b_0)}{1+(b/b_0)^2}\cos\chi,
\label{eq:dx_of_b}
\end{equation}
with
\begin{equation}
\Delta x_0 \equiv \frac{2 f_u}{1-\beta},
\qquad
f_u \equiv \frac{u}{v_1} = \sqrt{3 - 2\beta - C_J}.
\end{equation}
The probability distribution of $\Delta x$ is
\begin{equation}
\begin{split}
P(\Delta x)
&= \int_0^{2\pi} p(\chi)\,d\chi
\int_0^{b_{\max}}P(b) db\,
\delta\!\left(\Delta x - \Delta x(b, \chi)\right)\\
&=\frac{1}{\pi}
\int_{b_-}^{\min(b_+,\,b_{\max})}
\frac{P(b)\, db}{\sqrt{\Delta x(b)^2-\Delta x^2}},
\label{eq:Px_general}
\end{split} 
\end{equation}
with effective limits
\begin{equation}
R \equiv \frac{\Delta x_0}{|\Delta x|}>1,\qquad
b_\pm=b_0\Bigl(R\pm\sqrt{R^2-1}\Bigr).
\end{equation}
Here $b_{\max}$ denotes the maximum impact parameter over which the small-$b$ approximation to $P(b)$ is applied. In practice, we use $b_{\rm max}=2.5 \, r_{\rm Hill}$ for $i_2=0^{\circ}$ and $10^{\circ}$, and $b_{\rm max}=5 \, r_{\rm Hill}$ for $i_2=10^{\circ}$ for the analytic results shown in Fig \ref{fig:P_b_P_dx}.

In the coplanar limit, $P_{i_2=0}(b)$ is constant at small $b$. Within the Hill radius,
\begin{equation}\label{eq:Px_planar_integral}
P_{i_2=0}(\Delta x)
= \frac{C_0}{\pi}
\int_{b_-}^{\min(b_+,\,b_{\max})}
\frac{db}{\sqrt{\Delta x(b)^2-\Delta x^2}}.
\end{equation}
The transition between the small- and large-kick regimes occurs when the upper integration limit changes from $b_{\max}$ to $b_+$. This gives
\begin{equation}\label{eq:deltax_trans}
\Delta x_{\rm trans}\equiv \Delta x_0\frac{2b_0}{b_{\max}}
\quad\Longleftrightarrow\quad
b_+(\Delta x_{\rm trans})=b_{\max}.
\end{equation}
At small kicks $|\Delta x|\ll \Delta x_{\rm trans}$, the integral samples nearly the full allowed range of impact parameters and approaches a constant
\begin{equation}
P_{i_2=0}(\Delta x)\xrightarrow[|\Delta x|\to 0]{}
  \frac{C_0}{2\Delta x_0 b_0}.
\end{equation}
At large kicks $\Delta x_{\rm trans}\ll|\Delta x|<\Delta x_0$, only close encounters contribute. Using $\Delta x(b)\simeq 2\Delta x_0 b_0/b$ gives
\begin{equation}
P_{i_2=0}(\Delta x)
\simeq C_0\,\frac{2 b_0 \Delta x_0}{|\Delta x|^{2}}
\;\;\Rightarrow\;\;
P_{i_2=0}(\Delta x)\propto |\Delta x|^{-2}.
\end{equation}
Thus planar encounters produce a flat distribution at small $|\Delta x|$ and a power–law tail $P\propto|\Delta x|^{-2}$.

For a nonzero inclination, the small–$b$ distribution becomes linear. Substituting Equation \eqref{eq:P_b_inc} into Equation \eqref{eq:Px_general} gives
\begin{equation}
P_{i_2\neq 0}(\Delta x)
= \frac{C_0}{\pi a_1\sin{i_2}}
\int_{b_-}^{\min(b_+,\,b_{\max})}
\frac{b\, db}{\sqrt{\Delta x(b)^2-\Delta x^2}}.
\label{eq:Px_inclined_integral}
\end{equation}
For $|\Delta x|\ll\Delta x_{\rm trans}$, a flat plateau persists, though with a different normalization than the planar case:
\begin{equation}
P_{i_2\neq 0}(\Delta x)\xrightarrow[|\Delta x|\to 0]{}
\frac{C_0}{2 \Delta x_0 b_0 a_1\sin{i_2}}
\left(\frac{b_{\max}^3}{3}+b_0^2 b_{\max}\right).
\end{equation}
For $\Delta x_{\rm trans}\ll|\Delta x|<\Delta x_0$, the additional factor of $b$ in $p_b$ steepens the tail, giving
\begin{equation}
P_{i_2\neq 0}(\Delta x) \simeq \frac{C_0}{a_1\sin{i_2}}\frac{(2b_0\Delta x_0)^2}{|\Delta x|^3}
\;\;\Rightarrow\;\;
P_{i_2\neq 0}(\Delta x) \propto |\Delta x|^{-3}.
\end{equation}
Therefore inclination leaves the qualitative small-kick plateau intact, but changes the large-kick scaling (tail) from $|\Delta x|^{-2}$ to $|\Delta x|^{-3}$.

Both limits can be summarized by the unified expression
\begin{equation}\label{eq:p_dx_unif}
\begin{split}
&P(\Delta x)\\
&=\frac{C_0}{\pi}
\int_{b_-}^{\min(b_+,\,b_{\max})}
\frac{b}{\sqrt{(b^2+(a_1\sin{i_2})^2)(\Delta x(b)^2-\Delta x^2)}}\, db,
\end{split}
\end{equation}
with $|\Delta x|<\Delta x_0$, which interpolates smoothly between the planar and inclined limits.

\bibliography{citation}{}

\begin{thebibliography}{}
\expandafter\ifx\csname natexlab\endcsname\relax\def\natexlab#1{#1}\fi
\providecommand{\url}[1]{\href{#1}{#1}}
\providecommand{\dodoi}[1]{doi:~\href{http://doi.org/#1}{\nolinkurl{#1}}}
\providecommand{\doeprint}[1]{\href{http://ascl.net/#1}{\nolinkurl{http://ascl.net/#1}}}
\providecommand{\doarXiv}[1]{\href{https://arxiv.org/abs/#1}{\nolinkurl{https://arxiv.org/abs/#1}}}

\bibitem[{O. {Absil} {et~al.}(2013){Absil}, {Defr{\`e}re}, {Coud{\'e} du Foresto}, {Di Folco}, {M{\'e}rand}, {Augereau}, {Ertel}, {Hanot}, {Kervella}, {Mollier}, {Scott}, {Che}, {Monnier}, {Thureau}, {Tuthill}, {ten Brummelaar}, {McAlister}, {Sturmann}, {Sturmann}, \& {Turner}}]{2013A&A...555A.104A}
{Absil}, O., {Defr{\`e}re}, D., {Coud{\'e} du Foresto}, V., {et~al.} 2013, \bibinfo{title}{{A near-infrared interferometric survey of debris-disc stars. III. First statistics based on 42 stars observed with CHARA/FLUOR},} \aap, 555, A104, \dodoi{10.1051/0004-6361/201321673}

\bibitem[{O. {Absil} {et~al.}(2021){Absil}, {Marion}, {Ertel}, {Defr{\`e}re}, {Kennedy}, {Romagnolo}, {Le Bouquin}, {Christiaens}, {Milli}, {Bonsor}, {Olofsson}, {Su}, \& {Augereau}}]{2021A&A...651A..45A}
{Absil}, O., {Marion}, L., {Ertel}, S., {et~al.} 2021, \bibinfo{title}{{A near-infrared interferometric survey of debris-disk stars. VII. The hot-to-warm dust connection},} \aap, 651, A45, \dodoi{10.1051/0004-6361/202140561}

\bibitem[{P. {Arras} {et~al.}(2022){Arras}, {Wilson}, {Pryal}, \& {Baker}}]{2022ApJ...932...90A}
{Arras}, P., {Wilson}, M., {Pryal}, M., \& {Baker}, J. 2022, \bibinfo{title}{{Dust Accretion onto Exoplanets},} \apj, 932, 90, \dodoi{10.3847/1538-4357/ac625e}

\bibitem[{K. {Batygin} \& A.~C. {Petit}(2023){Batygin} \& {Petit}}]{2023ApJ...946L..11B}
{Batygin}, K., \& {Petit}, A.~C. 2023, \bibinfo{title}{{Dissipative Capture of Planets into First-order Mean-motion Resonances},} \apjl, 946, L11, \dodoi{10.3847/2041-8213/acc015}

\bibitem[{C. {Beauge} \& S. {Ferraz-Mello}(1994){Beauge} \& {Ferraz-Mello}}]{1994Icar..110..239B}
{Beauge}, C., \& {Ferraz-Mello}, S. 1994, \bibinfo{title}{{Capture in Exterior Mean-Motion Resonances Due to Poynting-Robertson Drag},} \icarus, 110, 239, \dodoi{10.1006/icar.1994.1119}

\bibitem[{J. {Binney} \& S. {Tremaine}(2008){Binney} \& {Tremaine}}]{2008gady.book.....B}
{Binney}, J., \& {Tremaine}, S. 2008, {Galactic Dynamics: Second Edition}

\bibitem[{A. {Bonsor} {et~al.}(2018){Bonsor}, {Wyatt}, {Kral}, {Kennedy}, {Shannon}, \& {Ertel}}]{2018MNRAS.480.5560B}
{Bonsor}, A., {Wyatt}, M.~C., {Kral}, Q., {et~al.} 2018, \bibinfo{title}{{Using warm dust to constrain unseen planets},} \mnras, 480, 5560, \dodoi{10.1093/mnras/sty2200}

\bibitem[{J.~A. {Burns} {et~al.}(1979){Burns}, {Lamy}, \& {Soter}}]{1979Icar...40....1B}
{Burns}, J.~A., {Lamy}, P.~L., \& {Soter}, S. 1979, \bibinfo{title}{{Radiation forces on small particles in the solar system},} \icarus, 40, 1, \dodoi{10.1016/0019-1035(79)90050-2}

\bibitem[{E. Chang {et~al.}(2026)Chang, Samra, Gao, \& Arras}]{Chang2026Dust}
Chang, E., Samra, D., Gao, P., \& Arras, P. 2026, \bibinfo{title}{Impact of Infalling Dust on Exoplanet Spectra,}

\bibitem[{J. {Chen} \& D. {Kipping}(2017){Chen} \& {Kipping}}]{2017ApJ...834...17C}
{Chen}, J., \& {Kipping}, D. 2017, \bibinfo{title}{{Probabilistic Forecasting of the Masses and Radii of Other Worlds},} \apj, 834, 17, \dodoi{10.3847/1538-4357/834/1/17}

\bibitem[{K.~M. {Deck} \& K. {Batygin}(2015){Deck} \& {Batygin}}]{2015ApJ...810..119D}
{Deck}, K.~M., \& {Batygin}, K. 2015, \bibinfo{title}{{Migration of Two Massive Planets into (and out of) First Order Mean Motion Resonances},} \apj, 810, 119, \dodoi{10.1088/0004-637X/810/2/119}

\bibitem[{B.~T. {Draine}(2011){Draine}}]{2011piim.book.....D}
{Draine}, B.~T. 2011, {Physics of the Interstellar and Intergalactic Medium}

\bibitem[{M. {Duncan} {et~al.}(1987){Duncan}, {Quinn}, \& {Tremaine}}]{1987AJ.....94.1330D}
{Duncan}, M., {Quinn}, T., \& {Tremaine}, S. 1987, \bibinfo{title}{{The Formation and Extent of the Solar System Comet Cloud},} \aj, 94, 1330, \dodoi{10.1086/114571}

\bibitem[{S. {Ertel} {et~al.}(2014){Ertel}, {Absil}, {Defr{\`e}re}, {Le Bouquin}, {Augereau}, {Marion}, {Blind}, {Bonsor}, {Bryden}, {Lebreton}, \& {Milli}}]{2014A&A...570A.128E}
{Ertel}, S., {Absil}, O., {Defr{\`e}re}, D., {et~al.} 2014, \bibinfo{title}{{A near-infrared interferometric survey of debris-disk stars. IV. An unbiased sample of 92 southern stars observed in H band with VLTI/PIONIER},} \aap, 570, A128, \dodoi{10.1051/0004-6361/201424438}

\bibitem[{E. {Everhart}(1968){Everhart}}]{1968AJ.....73.1039E}
{Everhart}, E. 1968, \bibinfo{title}{{Change in Total Energy of Comets Passing Through the Solar System},} \aj, 73, 1039, \dodoi{10.1086/110766}

\bibitem[{J.~A. {Fernandez}(1981){Fernandez}}]{1981A&A....96...26F}
{Fernandez}, J.~A. 1981, \bibinfo{title}{{New and evolved comets in the solar system},} \aap, 96, 26

\bibitem[{S. {Ferraz-Mello} \& M. {Sato}(1989){Ferraz-Mello} \& {Sato}}]{1989A&A...225..541F}
{Ferraz-Mello}, S., \& {Sato}, M. 1989, \bibinfo{title}{{The very-high-eccentricity asymmetric expansion of the disturbing function near resonances of any order.},} \aap, 225, 541

\bibitem[{T. {Gold}(1975){Gold}}]{1975Icar...25..489G}
{Gold}, T. 1975, \bibinfo{title}{{Resonant Orbits of Grains and the Formation of Satellites},} \icarus, 25, 489, \dodoi{10.1016/0019-1035(75)90016-0}

\bibitem[{P. {Goldreich} \& H.~E. {Schlichting}(2014){Goldreich} \& {Schlichting}}]{2014AJ....147...32G}
{Goldreich}, P., \& {Schlichting}, H.~E. 2014, \bibinfo{title}{{Overstable Librations can Account for the Paucity of Mean Motion Resonances among Exoplanet Pairs},} \aj, 147, 32, \dodoi{10.1088/0004-6256/147/2/32}

\bibitem[{H. {Goldstein} {et~al.}(2002){Goldstein}, {Poole}, \& {Safko}}]{2002clme.book.....G}
{Goldstein}, H., {Poole}, C., \& {Safko}, J. 2002, {Classical mechanics}

\bibitem[{M. {Henon} \& J.-M. {Petit}(1986){Henon} \& {Petit}}]{1986CeMec..38...67H}
{Henon}, M., \& {Petit}, J.-M. 1986, \bibinfo{title}{{Series Expansions for Encounter-Type Solutions of Hill's Problem},} Celestial Mechanics, 38, 67, \dodoi{10.1007/BF01234287}

\bibitem[{P. {Lavvas} \& T. {Koskinen}(2017){Lavvas} \& {Koskinen}}]{2017ApJ...847...32L}
{Lavvas}, P., \& {Koskinen}, T. 2017, \bibinfo{title}{{Aerosol Properties of the Atmospheres of Extrasolar Giant Planets},} \apj, 847, 32, \dodoi{10.3847/1538-4357/aa88ce}

\bibitem[{J. {Mang} {et~al.}(2022){Mang}, {Gao}, {Hood}, {Fortney}, {Batalha}, {Yu}, \& {de Pater}}]{2022ApJ...927..184M}
{Mang}, J., {Gao}, P., {Hood}, C.~E., {et~al.} 2022, \bibinfo{title}{{Microphysics of Water Clouds in the Atmospheres of Y Dwarfs and Temperate Giant Planets},} \apj, 927, 184, \dodoi{10.3847/1538-4357/ac51d3}

\bibitem[{J. {Mang} {et~al.}(2024){Mang}, {Morley}, {Robinson}, \& {Gao}}]{2024ApJ...974..190M}
{Mang}, J., {Morley}, C.~V., {Robinson}, T.~D., \& {Gao}, P. 2024, \bibinfo{title}{{Microphysical Prescriptions for Parameterized Water Cloud Formation on Ultra-cool Substellar Objects},} \apj, 974, 190, \dodoi{10.3847/1538-4357/ad6c4c}

\bibitem[{A. Meurer {et~al.}(2017)Meurer, Smith, Paprocki, {\v{C}}ert{\'i}k, Kirpichev, Rocklin, Kumar, Ivanov, Moore, Singh, Rathnayake, Vig, Granger, Muller, Bonazzi, Gupta, Vats, Johansson, Pedregosa, Curry, Saboo, Fernando, Kulal, Cimrman, \& Scopatz}]{Meurer2017SymPy}
Meurer, A., Smith, C.~P., Paprocki, M., {et~al.} 2017, \bibinfo{title}{SymPy: symbolic computing in Python,} PeerJ Computer Science, 3, e103, \dodoi{10.7717/peerj-cs.103}

\bibitem[{A. {Moro-Martin}(2013){Moro-Martin}}]{2013pss3.book..431M}
{Moro-Martin}, A. 2013, \bibinfo{title}{{Dusty Planetary Systems},} in Planets, Stars and Stellar Systems. Volume 3: Solar and Stellar Planetary Systems, ed. T.~D. {Oswalt}, L.~M. {French}, \& P.~{Kalas}, 431, \dodoi{10.1007/978-94-007-5606-9_9}

\bibitem[{C.~D. {Murray} \& S.~F. {Dermott}(1999){Murray} \& {Dermott}}]{1999ssd..book.....M}
{Murray}, C.~D., \& {Dermott}, S.~F. 1999, {Solar System Dynamics}, \dodoi{10.1017/CBO9781139174817}

\bibitem[{R.~A. {Murray-Clay} \& E.~I. {Chiang}(2005){Murray-Clay} \& {Chiang}}]{2005ApJ...619..623M}
{Murray-Clay}, R.~A., \& {Chiang}, E.~I. 2005, \bibinfo{title}{{A Signature of Planetary Migration: The Origin of Asymmetric Capture in the 2:1 Resonance},} \apj, 619, 623, \dodoi{10.1086/426425}

\bibitem[{E.~J. {Opik}(1976){Opik}}]{1976iecg.book.....O}
{Opik}, E.~J. 1976, {Interplanetary encounters : close-range gravitational interactions}

\bibitem[{H. {Rein} \& S.~F. {Liu}(2012){Rein} \& {Liu}}]{2012A&A...537A.128R}
{Rein}, H., \& {Liu}, S.~F. 2012, \bibinfo{title}{{REBOUND: an open-source multi-purpose N-body code for collisional dynamics},} \aap, 537, \href{https://ui.adsabs.harvard.edu/abs/2012A&A...537A.128R}{A128}, \dodoi{10.1051/0004-6361/201118085}

\bibitem[{A. {Shannon} {et~al.}(2015){Shannon}, {Mustill}, \& {Wyatt}}]{2015MNRAS.448..684S}
{Shannon}, A., {Mustill}, A.~J., \& {Wyatt}, M. 2015, \bibinfo{title}{{Capture and evolution of dust in planetary mean-motion resonances: a fast, semi-analytic method for generating resonantly trapped disc images},} \mnras, 448, 684, \dodoi{10.1093/mnras/stv045}

\bibitem[{M. {Sidlichovsky} \& D. {Nesvorny}(1994){Sidlichovsky} \& {Nesvorny}}]{1994A&A...289..972S}
{Sidlichovsky}, M., \& {Nesvorny}, D. 1994, \bibinfo{title}{{Temporary capture of grains in exterior resonances with the Earth: planar circular restricted three-body problem with Poynting-Robertson drag.},} \aap, 289, 972

\bibitem[{D. {Tamayo} {et~al.}(2020){Tamayo}, {Rein}, {Shi}, \& {Hernandez}}]{2020MNRAS.491.2885T}
{Tamayo}, D., {Rein}, H., {Shi}, P., \& {Hernandez}, D.~M. 2020, \bibinfo{title}{{REBOUNDx: a library for adding conservative and dissipative forces to otherwise symplectic N-body integrations},} \mnras, 491, 2885, \dodoi{10.1093/mnras/stz2870}

\bibitem[{P. Virtanen {et~al.}(2020)Virtanen, Gommers, Oliphant, Haberland, Reddy, Cournapeau, Burovski, Peterson, Weckesser, Bright, {van der Walt}, Brett, Wilson, Millman, Mayorov, Nelson, Jones, Kern, Larson, Carey, Polat, Feng, Moore, {VanderPlas}, Laxalde, Perktold, Cimrman, Henriksen, Quintero, Harris, Archibald, Ribeiro, Pedregosa, {van Mulbregt}, \& {SciPy 1.0 Contributors}}]{2020SciPy-NMeth}
Virtanen, P., Gommers, R., Oliphant, T.~E., {et~al.} 2020, \bibinfo{title}{{{SciPy} 1.0: Fundamental Algorithms for Scientific Computing in Python},} Nature Methods, 17, 261, \dodoi{10.1038/s41592-019-0686-2}

\bibitem[{S.~J. {Weidenschilling} \& A.~A. {Jackson}(1993){Weidenschilling} \& {Jackson}}]{1993Icar..104..244W}
{Weidenschilling}, S.~J., \& {Jackson}, A.~A. 1993, \bibinfo{title}{{Orbital Resonances and Poynting-Robertson Drag},} \icarus, 104, 244, \dodoi{10.1006/icar.1993.1099}

\bibitem[{S.~P. {Wyatt} \& F.~L. {Whipple}(1950){Wyatt} \& {Whipple}}]{1950ApJ...111..134W}
{Wyatt}, S.~P., \& {Whipple}, F.~L. 1950, \bibinfo{title}{{The Poynting-Robertson effect on meteor orbits},} \apj, 111, 134, \dodoi{10.1086/145244}

\bibitem[{S. {Yabushita}(1980){Yabushita}}]{1980A&A....85...77Y}
{Yabushita}, S. 1980, \bibinfo{title}{{On exact solutions of diffusion equation in cometary dynamics},} \aap, 85, 77

\end{thebibliography}
\bibliographystyle{aasjournalv7}

\end{document}